# Design of Reconfigurable Multi-Operand Adder for Massively Parallel Processing

**Shilpa Mayannavar**[1] and **Uday Wali**[2]

[1]Dept. ECE, SGBIT, Belagavi - 590010
Email: mayannavar.shilpa@gmail.com

[2]C-Quad Research,
SGBIT, Belagavi - 590010
Email: udaywali@gmail.com

**ABSTRACT**

*The paper presents a systematic study and implementation of a reconfigurable combinatorial multi-operand adder for use in Deep Learning systems. The size of carry changes with the number of operands and hence a reliable algorithm to estimate exact number of carry bits is needed for optimal implementation of a reconfigurable multi-operand adder. A combinatorial multi-operand adder can be faster compared to a sequential implementation using a two operand adder. Use cases for such adders occur in modern processors for deep neural networks. Such processors require massively parallel computing resources on chip. This paper presents a method to estimate the upper bound on the size of carry. A method to compute the exact number of carry bits required for a multi-operand addition operation. A fast combinatorial parallel 4-operand adder module is presented. An algorithm to reconfigure these adder modules to implement larger adders is also described. Further, the paper presents two compact but slower iterative structures that implement multi-operand addition, iterating with one column at a time till the entire word is covered. Such serial/iterative operations are slow but occupy small space while parallel operations are fast but use large silicon area on chip. Interestingly, the area-to-throughput ratio of two architectures can tilt in favor of slower, smaller and large number units instead of the fewer numbers of fast and large compute units. A lemma presented in the paper may be used to identify the condition when such tilt occurs. Potentially, this can save silicon space and increase the throughput of chips for high performance computing. Simulation results of a 16 operand adder and using an set of 4-operand adders for use in neural networks have been presented. Simulation results show that performance gain improves as the number of operations or operands increases.*

## 1. INTRODUCTION

Artificial Intelligence (AI) and Deep Learning Neural Networks (DLNNs) have become part of everyday computing. Convolutional Neural Networks (CNNs) used in image classification and recognition are a typical example of such deep learning systems [2012 Krizhevsky]. Recently announced GPT3, Natural Language Processing application uses hundreds of billions of words to train a neural network [2020 OpenAI, 2020 Brown]. The network has 96 Attention layers and 175 billion trainable parameters with a batch size of 3.2 million tokens. Currently GPUs are being used to address computational demands of such High Performance Computing (HPC) needs. Modern GPUs like NVidia A100 bring the computing power of super computers to servers [2020 Nvidia] for data center applications. A100 processor can support up to 1500GB/s memory transfer rate and may be reconfigured to function as upto-7 instances of high performance GPU accelerators. Each of such GPU instances can be configured to meet application specific resources available on this single chip CPU. Several processors like Google TPU [2017 Jouppi], Intel Nervana [2017 Koster] and IBM TrueNorth [2015 DeBole] were designed with AI centric architectures, moving away from GPU based designs. Commercial processors are also extending their capabilities, e.g., Intel AVX-512 extensions added four new instructions called Vector Neural Network Instructions (VNNI) to ease CNN implementations [2014 Intel]. These new instructions support 'multiply and accumulate' operations that can be concatenated to implement multi-operand operations. Abdelouahab et al. have discussed some of the challenges in the design of multi-operand adders for CNNs and also reported the problems faced during the implementation [2018 Abdelouahab]. Some implementations were reported in the early '70s but as there was no justification for a multi-operand adder, the work did not attract further development. The problem of adding four or more operands has been discussed by Singh et al. [1973 Singh] wherein a method of bit-partitioning is used to obtain the sum in m+1 clock cycles when the operands are m-bits. Design of multi-operand binary adder using two methods viz., a tree of 2-input 1-output adders and a tree of 3-input 2-output carry save adders is discussed by Atkins and Ong [1978 Atkins]. Tajasob et al. have proposed multi-bit adder using approximate computing methods. No theoretical work in support of multi-operand adders has been reported [2018 Tajasob].

While the current processor designs support module level reconfiguration, there is possibility of designing individual modules that can be reconfigured internally to enhance module performance. In this paper, a reconfigurable multi-operand adder is discussed. The need for fast multi-operand addition is encountered frequently at the output of ANN nodes (neurons). The output of a typical neuron is calculated as weighted sum of N-inputs followed by a non-linear activation function. The number of inputs to a neuron can range from few to few thousand. For example, ImageNet [2012 Krizhevsky] uses 650,000 neurons in five convolution layers and three fully connected layers. The input size of each convolution layers range from 363 to 2304. Another DLNN, LeNet5 uses neurons with up to 4704 inputs [1998 LeCun]. Multi-operand addition will be a computationally significant operation in many such DLNNs. It is also easy to see that the number of operands will vary as the computations shift from one layer to the next. In several models implementing DLNNs, inputs are scaled by a weight factor and then summed to compute the output, e.g. Multi-Layer Perceptron (MLP). Such operations are combined in to a Streamed Multiply and Accumulate (sMAC) instruction as shown in Figure 1 or implemented in hardware as a systolic array using cells for each operation, and gradually combining the results [1981 Kung]. However, this will require at least (N-1) operations for N-operands. Using a tree of adders can increase the speed to $\log_2(N)$ if N/2 adders are available, but the number of operations will remain same. Therefore, using conventional two operand adders in such scenario may not be very efficient or convenient. Use of a reconfigurable, combinatorial multi-operand addition can reduce the overall time and area, as shown in this paper.

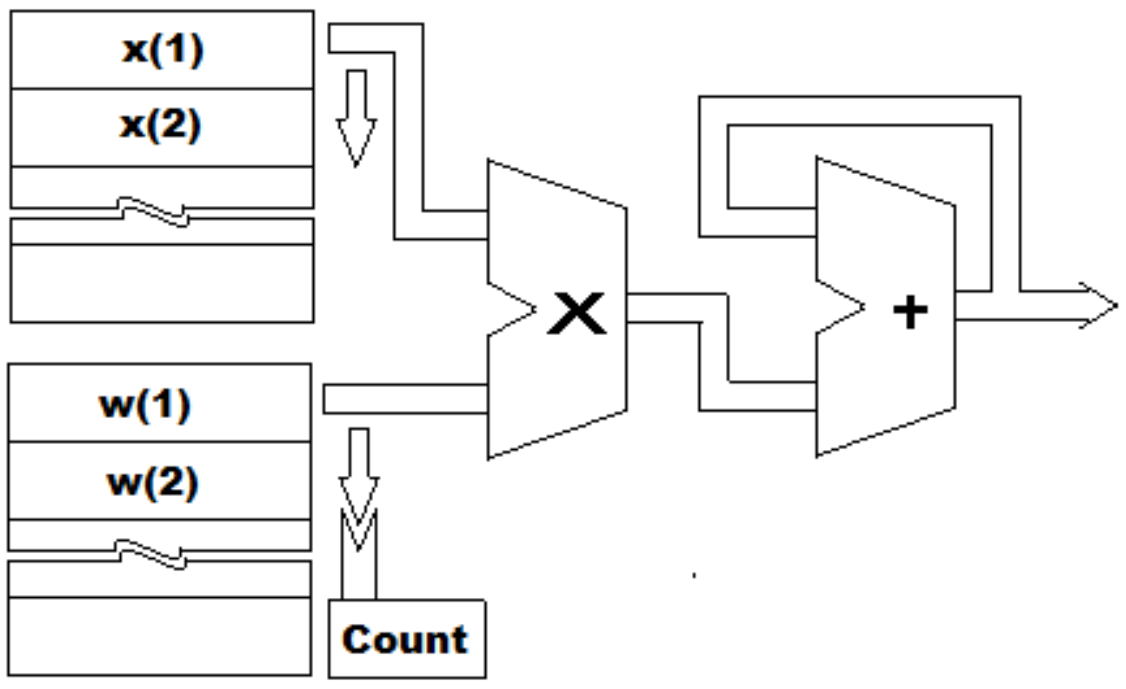

**Figure 1.** Implementing a Streamed MAC Instruction

Multi-operand addition differs from two operand addition in the number of carry bits required to complete the operation. In order to address this issue, systematic study of multi-operand adder is required. Estimation of the upper bound on the carry for arbitrary number of operands is the first step. Number of bits required to accommodate this upper bound can be calculated easily. Some initial results in this direction were reported by the authors in [2019 Mayannavar]. Further discussions and necessary derivations are presented in this paper.

Multi-operand addition is generally handled as an iterative procedure or data flow structure, considering two operands at a time. An alternate option is to perform multi-operand addition as a single integrated step. However, there is very little published literature in this area. One of the difficulties in implementing a multi-operand addition is to know the exact number of carry bits. Using more number of bits slows down the computation while using smaller number of bits can produce overflows. A lemma to quickly compute the number of bits is presented in this paper.

Section 2 contains the theoretical background required for implementing a multi-operand addition. These observations are used to estimate the maximum number of carry and sum bits required for N-operand addition. Section 3 describes algorithms for multi-operand addition, based on the discussions made in section 2. Section 4 and 5 contain design of a 4-operand M-bit serial and parallel adder modules respectively. Typically, given an algorithm, it is possible to implement it using serial or parallel hardware. In a conventional computing environment, serial computation is characterized by lower throughput and less hardware. On the other hand parallel computation is fast but involves larger chip area. Computational load of a DLNN involves large vector/tensor operations. Therefore, there is a case of trade-off between serial and parallel implementations in terms of area requirements vs throughput. Performance comparison of serial and parallel modules in massively parallel environment is discussed in section 6. Construction of a 16-operand adder using the previously described 4-operand module is presented in section 7. Section 8 describes the use of multi-operand adders in implementing an N-input neuron considering two types of neural architectures as an example. Section 9 gives simulation results of multi-operand adders that confirm the theory and designs discussed in its preceding sections. Comparison of proposed multi-operand adder with the conventional Carry Look Ahead (CLA) adder is presented in section 10. Conclusion is provided at the end.

## 2. THEORETICAL CONSTRUCTION FOR MULTI-OPERAND ADDITION

Following discussions can be applied to any number base (binary, decimal, octal, etc), for any number of columns (bits, digits, octets etc) and for any number of operands.

### 2.1 Basic Notations

Let $N$ be the number of operands, $M$ be the number of columns (bits, digits, etc.), $k$ be the number base, $Z$ be the total sum, $C$ be the value of a carry and $S$ be the column sum. The total sum $Z$ can be expressed as, $Z = kC + S$, where $0 \leq S < k$. $S$ may be expressed as

$$S = Z \, mod \, k \tag{1}$$

From eqn. (1), $if(Z - S) > 0$, it is a multiple of $k$. Or

$$kC = (Z - S) \tag{2}$$

For 2-operand addition, maximum value of $C$ is $2_k$-1, which will be numerically equal to 1 for all $k > 1$.

In case of multi-operand addition, $C$ can be more than $(k - 1)$, extending to more than one column. Therefore, as a general form, $C$ may be expressed as

$$C = \sum_{i=0}^{p} c_i \, k^i \text{ , where } 0 \leq c_i < k \tag{3}$$

In eqn. (1), when $Z$ is a multiple of $k$, we get, $S = 0$ and therefore, $Z$ can be expressed completely in terms of $C$. Number of rows ($N$, is independent of $k$ but without loss of any generality, it may also be expressed as a number in base $k$, similar to $C$, as

$$N = \sum_{i=0}^{p} n_i \, k^i \text{ , } where \, 0 \leq n_i < k \tag{4}$$

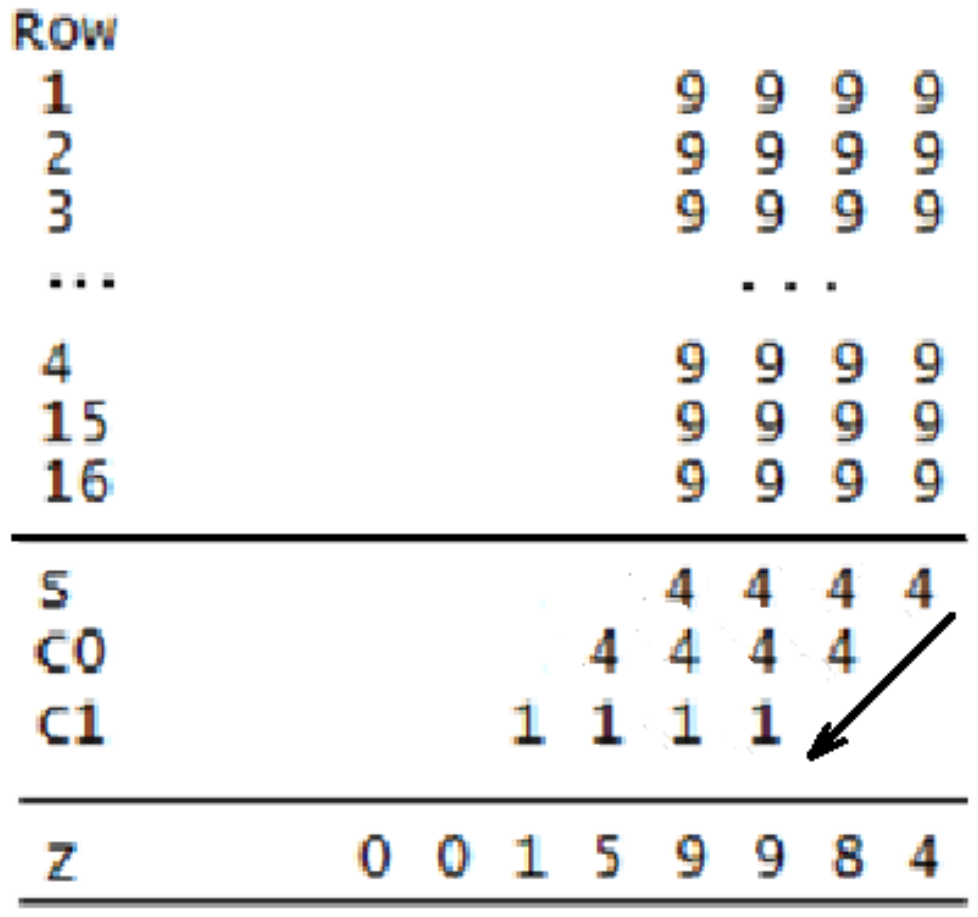

**Figure 2.** An illustrative example for Multi-operand addition in base 10 (N=16, M=4, k=10)

An illustrative example for 16-operand, 4-digit addition in base 10 (N=16, M=4, k=10) is shown in Figure 2. Initially, we will study a single column addition and then extend the results to multiple columns. For any base $k$, single column addition, the maximum value of an operand is $(k - 1)$. Therefore, for $N$-operand addition, the total sum is $Z = N(k - 1)$. This is also the condition for maximum carry. In the following discussion, base of a number is assumed to be $k$ or written as $(k)$ where necessary.

***Lemma 1:*** *For two-operand one-column addition $N = 2$, $M = 1$ and for a base of $k$, maximum carry is $C_{(k)} = 1$ and column sum is $S_{(k)} = k - 2$.*

***Proof:*** Total sum when all the operands are $(k-1)$ is

$$Z = N(k-1) = 2(k-1)$$

where, $S$ and $C$ may be calculated using eqn. (1) and eqn. (2) as,

$$S = Zmodk = (-2)modk = k - 2, \forall k \geq 2$$

$$kC = \{2k - 2 - k + 2\} = k, \therefore C = 1 \qquad \text{QED}$$

This result holds for all $k$. This can be verified easily in trivial cases, as below;

| | | | |
|---|---|---|---|
| $k = 2$: | $Z = 10_{(2)}$, | $C = 1_{(2)}$, | $S = 2 - 2 = 0_{(2)}$; |
| $k = 8$: | $Z = 16_{(8)}$, | $C = 1_{(8)}$, | $S = 8 - 2 = 6_{(8)}$; |
| $k = 10$: | $Z = 18_{(10)}$, | $C = 1_{(10,)}$, | $S = 10 - 2 = 8_{(10,)}$; |
| $k = 16$: | $Z = 1E_{(16)}$, | $C = 1_{(16)}$, | $S = 16 - 2 = E_{(16)}$; |

As we add more rows to this basic operation, $(k-1)$ terms are added to $Z$, $S$ will be decreased by 1 and $C$ will be increased by 1, for every additional row. Interestingly, when $odk = 0, S = 0$, adding next row adds another $k - 1$ term, which is absorbed completely by $S$ and therefore $C$ remains unchanged. This happens whenever the number of rows is a multiple of $k$. We can state this as follows:

***Lemma 2:*** *For maximum carry condition, as $N$ increases by 1, $S$ decreases by 1 and $C$ increases by 1, except when $N = nk + 1$.*

**Proof:** For a given $N$ rows each containing a number $(k-1)$, let $Z$ be the sum of $N$ rows, which has numeric value

$$Z_N = kC_N + S_N \tag{5}$$

where the suffix $(\cdot)_N$ represents the values when $N$ rows are considered. Increasing the number of rows from $N$ to $N + 1$, we get

$$Z_{N+1} = Z_N + (k-1)$$

$$= kC_N + S_N + k - 1$$

Or

$$Z_{N+1} = k\,(C_N + 1) + (S_N - 1)$$

$$= kC_{N+1} + S_{N+1} \tag{6}$$

when $Z_N = nk$, we get $S_N = 0$. Adding one more row will not result in addition of carry $C$ because

$$S_{N+1} = (S_N - 1)\,modk = -1\,mod\,k$$

$$= (k-1)$$

$S_{N+1} < k$ and therefore, additional row value will be absorbed into $S$ and hence there is no increment to $C$. QED

Consider the following numeric examples of one-column addition:

| | | | | |
|---|---|---|---|---|
| $k = 2$: | $N = 3$: | $Z = 1\,1_{(2)}$ | $C = 1_{(2)}$ | $S = 1_{(2)}$ |
| | $N = 4$: | $Z = 10\,0_{(2)}$ | $C = 10_{(2)}$ | $S = 0_{(2)}$ |
| | $N = 5$: | $Z = 10\,1_{(2)}$ | $C = 10_{(2)}$ | $S = 1_{(2)}$ |
| | $N = 6$: | $Z = 11\,0_{(2)}$ | $C = 11_{(2)}$ | $S = 0_{(2)}$ |
| $k = 10$: | $N = 19$: | $Z = 17\,1_{(10)}$ | $C = 17_{(10)}$ | $S = 1_{(10)}$ |
| | $N = 20$: | $Z = 18\,0_{(10)}$ | $C = 18_{(10)}$ | $S = 0_{(10)}$ |
| | $N = 21$: | $Z = 18\,9_{(10)}$ | $C = 18_{(10)}$ | $S = 9_{(10)}$ |

As we add rows, $N \; mod \; k$ iterates through 0 to $k - 1$ and $C$ misses one count per iteration. As $N$ increases, difference between $C$ and $N$ increases, making $C$ smaller by 1 for every $k$ rows.

In all the cases, the carry $C$ will not exceed $N - 1$. This indeed is an upper bound on the carry, and can be stated as a theorem as described below:

***Theorem:*** *An upper bound on value of the carry is numerically equal to the number of operands minus one, irrespective of the number of columns or the number system used i.e., if there are $N$ operands, the upper bound on the value of carry is $N - 1$.*

***Proof:*** For $\mathrm{M} = 1$,

$$\mathrm{S} = \mathrm{N(k-1) \; mod \; k}$$

$$\mathrm{kC = N(k-1) - (N(k-1) \; mod \; k)}$$

Or
$$\mathrm{kC = N(k-1) - (-N \; mod \; k)} \tag{7}$$

Let us consider specific cases of $N$ in relation to $k$:

For $\mathrm{N} < k$, $\mathrm{(-N \; mod \; k) = (k - N)}$

$$\therefore \mathrm{kC = N(k-1) - (k-N)}$$

or
$$\mathrm{C = N - 1} \tag{8}$$

For $\mathrm{N = nk}$, $\mathrm{(-N \; modk) = 0}$

$$\therefore \mathrm{kC = nk(k-1) - 0}$$

or
$$\mathrm{C = N - n} \tag{9}$$

which is less than $N - 1$.

For $\mathrm{N} > k$, N can be expressed as

$$\mathrm{N = nk + r} \tag{10}$$

$$\mathrm{(-N \; modk) = (-r \; mod \; k) = k - r}$$

$$\therefore \mathrm{kC = (nk + r)(k-1) - (k-r)}$$

$$\mathrm{kC = nk^2 - (n - r + 1)k}$$

Eliminating $k$ and combining with eqn. (10), we get

$$C = N - 1 - n \tag{11}$$

which is less than $N - 1$. Therefore, from eqn. (8), eqn. (9) and eqn. (11), upper bound on $C$ for all values of $N$ is $N - 1$. Also notice that $C$ will gradually move away from $N - 1$ as $n$ increases. We will use induction to prove that the above theorem holds for any column. Let $Z_m$ be the column sum including carry. It can be expressed as

$$Z_m = N\,(k-1) + C_{m-1} \tag{12}$$

where $C_{m-1}$ represents carry from previous column. Let us assume that the carry from previous column has an upper bound of $N - 1$. Substituting the value of carry from previous column, we can write

$$Z_m = N(k-1) + N - 1$$

$$= Nk - 1 \tag{13}$$

$$= -1 \; mod \; k$$

$$= k - 1 \tag{14}$$

The carry can now be written as

$$kC_m = Nk - 1 - (k-1)$$

$$\therefore \; C_m = N - 1 \tag{15}$$

Therefore, by induction, upper bound on the value of carry for any multi-operand addition is $\mathrm{N} - 1$.

QED

It may be noted that $N - 1$ is a generalized upper bound. It is possible to derive a tighter bound in specific cases like for $N > k, N = nk$, etc. As $N$ increases, the upper bound reduces to compensate for the $S$ term becoming zero for every $k$-th row. Before proceeding with these special cases, let us see some examples of multi-operand addition for different values of $N$ and $k$ as shown in Table 1 and Table 2. Note that, $N < k$ is valid only for $k > 2$, as $N < 2$ is not an addition at all (See Table 1).

### 2.1 Number of Carry Columns

Upper bound on $C$ is an indication of number of columns required to perform multi-column addition. Often, the number of columns is more useful than the upper bound on carry. From eqn. (3) it is easy to see that the number of columns increases by one for every increase in $p$. The actual shift occurs when $N$ is somewhat higher than $k^p$.

***Corollary:*** *The number of columns required to represent the carry is* $log_k(N-1)$.

This is an obvious result because the maximum value of carry is $N - 1$ by the above theorem. This corollary can also be derived from an alternate observation, as follows.

Result of an N-operand M-column addition can be expressed as

$$Z = N\,\{\,(k-1)\,k^{M-1} + (k-1)k^{M-2} + \ldots + (k-1)\} \tag{16}$$

Or

$$Z = N\,(k^M - 1) \tag{17}$$

**Table 1.** Table showing upper bound on the value of carry for 1-column addition a) for N<k b) For N>k c) For N=nk

| | k | N | M | Z | | n | $C_{Actual}$ | | $C_{UB}$ |
|---|---|---|---|---|---|---|---|---|---|
| | | | | C | S | | $C_{(k)}$ | C | N-1 |
| For N < k | 10 | 2 | 1 | 1 | 8 | 0 | 1 | 1 | 1 |
| | | 4 | 1 | 3 | 6 | 0 | 3 | 3 | 3 |
| | 16 | 10 | 1 | 9 | 6 | 0 | 9 | 9 | 9 |
| | | 15 | 1 | E | 1 | 0 | E | 14 | 14 |

(a)

| | k | N | M | Z | | n | $C_{Actual}$ | | $C_{UB}$ |
|---|---|---|---|---|---|---|---|---|---|
| | | | | C | S | | $C_{(k)}$ | C | N-1-n |
| For N > k | 2 | 5 | 1 | 10 | 1 | 2 | 10 | 2 | 2 |
| | | 7 | 1 | 11 | 1 | 3 | 11 | 3 | 3 |
| | 10 | 11 | 1 | 9 | 9 | 1 | 9 | 9 | 9 |
| | | 18 | 1 | 16 | 2 | 1 | 16 | 16 | 16 |
| | 16 | 20 | 1 | 12 | C | 1 | 12 | 18 | 18 |
| | | 33 | 1 | 1E | F | 2 | 1E | 30 | 30 |

(b)

| | k | N | M | Z | | n | $C_{Actual}$ | | $C_{UB}$ |
|---|---|---|---|---|---|---|---|---|---|
| | | | | C | S | | $C_{(k)}$ | C | N-n |
| For N=nk | 2 | 4 | 1 | 10 | 0 | 2 | 10 | 2 | 2 |
| | | 12 | 1 | 110 | 0 | 6 | 110 | 6 | 6 |
| | 10 | 20 | 1 | 18 | 0 | 2 | 18 | 18 | 18 |
| | | 50 | 1 | 45 | 0 | 5 | 45 | 45 | 45 |
| | 16 | 16 | 1 | F | 0 | 1 | F | 15 | 15 |
| | | 48 | 1 | 2D | 0 | 3 | 2D | 45 | 45 |

(c)

**Table. 2.** Table showing Upper bound on the value of carry for multi-column addition

| k | N | M | Z | | $C_{Actual}$ | | $C_{UB}$ |
|---|---|---|---|---|---|---|---|
| | | | C | S | $C_{(k)}$ | C | N-1 |
| 2 | 2 | 3 | 1 | 110 | 1 | 1 | 1 |
| | 4 | 3 | 11 | 100 | 11 | 3 | 3 |
| | 7 | 3 | 110 | 001 | 110 | 6 | 6 |
| | 7 | 5 | 110 | 11001 | 110 | 6 | 6 |
| | 10 | 3 | 1000 | 110 | 1000 | 8 | 9 |
| | 64 | 3 | 111000 | 000 | 111000 | 56 | 63 |
| 10 | 2 | 3 | 1 | 998 | 1 | 1 | 1 |
| | 4 | 3 | 3 | 996 | 3 | 3 | 3 |
| | 10 | 3 | 9 | 990 | 9 | 9 | 9 |
| | 15 | 4 | 14 | 9985 | 14 | 14 | 14 |
| | 1112 | 3 | 1110 | 888 | 1110 | 1110 | 1111 |
| 16 | 2 | 3 | 1 | FFE | 1 | 1 | 1 |
| | 4 | 3 | 3 | FFC | 3 | 3 | 3 |
| | 18 | 3 | 11 | FEE | 11 | 17 | 17 |
| | 65520 | 2 | FEF0 | 10 | FEF0 | 65264 | 65519 |

The term $(\mathrm{k}^{\mathrm{M}} - 1)$ represents the largest number that can be represented using $\mathrm{M}$ columns. Therefore, for all $\mathrm{p} \geq 1$ and $\mathrm{N}$ in the range $\mathrm{k}^{\mathrm{p}} > N \geq \mathrm{k}^{\mathrm{p}-1}, \mathrm{p}$ columns are sufficient to represent the carry. This is a much higher bound than $\log_{\mathrm{k}}(\mathrm{N} - 1)$, but both will require the same number of columns. E.g., for $\mathrm{k} = 10$, $\mathrm{p} = 2$, $100 > N \geq 10$ , two columns are sufficient to represent a carry, but the value of carry will have an upper bound of $\mathrm{N} - 1$ as observed earlier.

It may be noted that the difference between the actual maximum carry and the above defined upper bound increases as $N$ increases because of the effect discussed earlier. In the next section, a much better estimate of the bound is derived.

### 2.2 Column Transition

In eqn. (4), i.e., $N = \sum_{i=0}^{p} n_i\, k^i$, $p$ is meaningful only when $n_p \neq 0$, i.e., $0 < n_{\mathrm{p}} < k$ and $0 \leq \mathrm{n_i} < k,\ i = 0 \ldots \mathrm{p} - 1$. When increasing the number of rows from $N = (k-1)\sum k^{p-1}$ to $N = k^p$, upper bound on number of columns required for $Z$ will increase by one column. However, the actual number of columns will not increase immediately for $N = k^p$, but slightly higher. It is interesting to study when the number of columns actually increases after $k^p$ boundary.

From eqn. (17), for any $N$ in the range $2 \leq \mathrm{N} < k$, $Z$ will require just one more column. For $N = k$, eqn. (17) becomes

$$Z = k(k^M - 1) = (k^{M+1} - k)$$

which is much less than $(k^{M+1} - 1)$ and therefore $M + 1$ columns are sufficient to represent $Z$. When $N = k + 1$,

$$Z = (k^{M+1} - 1) + (k^M - k)$$

The term $(k^M - k) > 1$ for all $M > 1$, or, $Z > k^{M+1}$. Therefore we will require $M + 2$ columns to represent $Z$, i.e., the number of columns increases by one. Similar increment is expected at every increase in order of $N$, i.e., $k^2$, $k^3$ etc. However, multiplication by $k^p$ shifts the $(k^M - 1)$ by $p$ columns, leaving room for few more rows of $(k^M - 1)$ terms. Especially when $p > M$, $N$ can increase substantially from $k^p$ without requiring additional columns.

Combining eqn. (17) and eqn. (4) we can write

$$Z = \left(\sum_{i=0}^{p} n_i\, k^i\right)(k^M - 1)$$

$$= n_p k^{M+p} - n_p k^p + \left(\sum_{i=0}^{p-1} n_i\, k^i\right)(k^M - 1) \tag{18}$$

Changing $N$ from $(k^{M+p} - 1)$ to $k^{M+p}$ will increase the number of columns by 1. In eqn. (18) this occurs if $n_p = 1$ in $n_p k^{M+p}$ and all other terms are zero. However, existence of $-n_p k^p$ term in the eqn. (18) delays this slightly beyond $k^p$. At the transition, $n_p = 1$ (from observation 2) we can rewrite eqn. (18) as

$$Z = k^{M+p} - k^p + \left(\sum_{i=0}^{p-1} n_i\, k^i\right)(k^M - 1) \tag{19}$$

The transition occurs when

$$-k^p + \left(\sum_{i=0}^{p-1} n_i\, k^i\right)(k^M - 1) \geq 0$$

Or when

$$\left(\sum_{i=0}^{p-1} n_i\, k^i\right)(k^M - 1) \geq k^p \tag{20}$$

So, the problem of finding the *N*, when the number of carry bits increases can be stated as the problem of finding $n_i, i = 0..p-1$ such that eqn. (20) is satisfied. Note that $n_p = 1$.

A numerical example for column transition is shown in Table 3. For $k$=2, $N$=15, $p$=3 and number of carry bits is 4, when $N$ increases by 1 i.e., $N$=16, we have $p$=4 and number of carry bits is expected to be 5. However, the transition is delayed till $N$ increases such that eqn. (20) is satisfied. Minimum coefficients that satisfy the equation $(n_3k^3+n_2k^2+n_1k^1+n_0k^0)(k^3-1) \geq k^4$ are

$$n_3=0,\ n_2=0,\ n_1=1 \text{ and } n_0=1 \text{ or } 0011_{(2)} = 3_{(10)}$$

Therefore transition occurs when $N$ shifts by 3, i.e., N=16+3=19. This can be seen easily in Table 3. Also notice from Table 3 that when N=16, S=000(2), $(k^M-1)$ shifts by 4 columns to the left, with $S$ taking three zeros and $C$ taking the remaining one zero.

**Table 3.** Example for Column Transition

| k | M | $k^M$-1 | N | $N_{(k)}$ | p | $k^p$ | $Z_{(k)}$ | | $Z_{(10)}$ |
|---|---|---|---|---|---|---|---|---|---|
| | | | | | | | C | S | |
| 2 | 3 | 111 | 15 | 1111 | 3 | 1000 | 1101 | 001 | 105 |
| | 3 | 111 | 16 | 10000 | 4 | 10000 | 1110 | 000 | 112 |
| | 3 | 111 | 19 | 10011 | 4 | 10000 | 10000 | 101 | 133 |

## 3. ALGORITHMS FOR MULTI-OPERAND BINARY ADDITION

### 3.1 Design of look-up-table (LUT)

We will use binary numbers $(k = 2)$ in the following sections. For a 4-operand, 1-column addition, the upper bound on $C$ is 3, therefore, $C$ has only four possible values $(00{:}11)$ and the total sum, $Z$ has only 5 possible values $(000{:}100)$. Therefore, 2-bits are sufficient to represent the carry of 4-operand addition and 3-bits would be sufficient to represent column sum (including carry for column). From corollary, $M + p$ bits are sufficient to compute *N*-operand addition for *M*-bit data.

Column sum for input data bits can be implemented as a look-up table of $N * (1 + p)$ bits or as a hardwired 1's count logic. As $N$ grows, size of the look-up table increases exponentially and therefore, the lookup table has to be limited to a small $N$.

| Input | Output |
|---|---|
| 0 0 0 0 | 0 0 0 |
| 0 0 0 1 | 0 0 1 |
| 0 0 1 0 | 0 0 1 |
| 0 1 0 0 | 0 0 1 |
| 1 0 0 0 | 0 0 1 |
| 0 0 1 1 | 0 1 0 |
| 0 1 1 0 | 0 1 0 |
| 1 1 0 0 | 0 1 0 |
| 1 0 0 1 | 0 1 0 |
| 0 1 0 1 | 0 1 0 |
| 1 0 1 0 | 0 1 0 |
| 0 1 1 1 | 0 1 1 |
| 1 1 1 0 | 0 1 1 |
| 1 1 0 1 | 0 1 1 |
| 1 0 1 1 | 0 1 1 |
| 1 1 1 1 | 1 0 0 |

**Figure 3.** Structure of 4x3 LUT. (The table is split horizontally to improve layout.)

The I/O map of $4x3$ LUT (4-bit input, 3-bit output) is shown in Figure 3. It is possible to build a RAM based LUT given in Figure 3 or build an optimized combinatorial circuit (one's count logic) as shown in Figure 4.

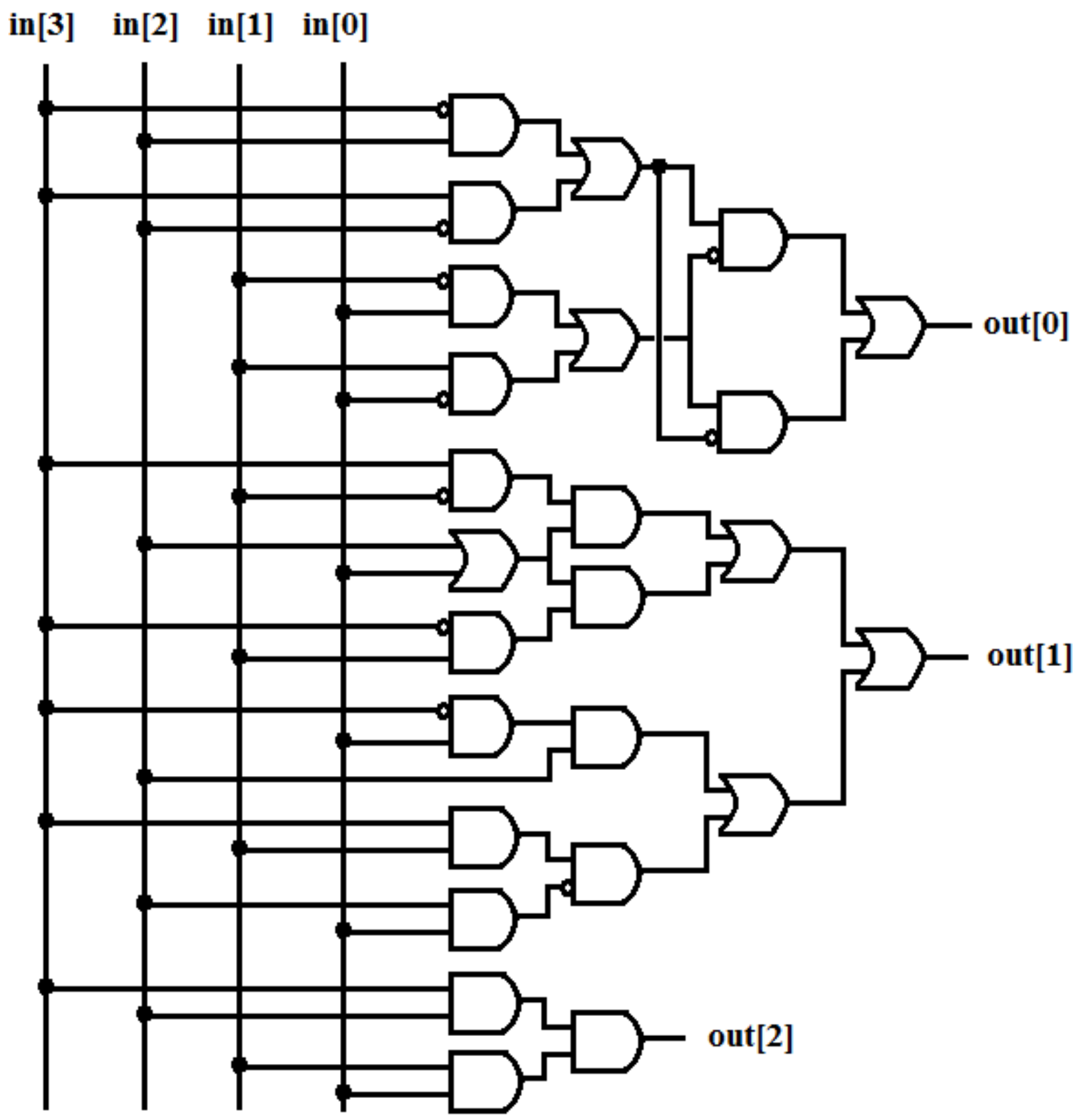

**Figure 4.** One's count logic equivalent to 4x3 LUT

The advantage of using this circuit instead of memory based LUT is that the adder can be implemented without any need for clocking. This implementation reduces overall time for computation and would consume a minimum silicon area as compared to a memory-like implementation. Longest path in Figure 4 has four gates (two input). So the speed performance of the LUT is good.

### 3.2 Proposed Algorithms for Multi-operand Serial addition

Following algorithms implement column wise addition propagating the carry towards left side (higher) columns as shown in Figure 5. Two algorithms discussed below differ in the way they handle the carry.

***Algorithm-1:***

This algorithm is similar to hand calculation. The algorithm is as follows:

a) Starting from LSB, each $i$-th column will have $N$ data rows and $i-1$ carry rows to be added.

b) The partial sum is copied to carry buffer, starting at column $i$ and extending to the left.

c) Move to next column to the left and repeat till all $M+p$ columns are computed.

d) At the end of the iteration, result is available in output buffer.

It is important to note that the value of carry in every column is in the range $0 \leq c_i < k$.

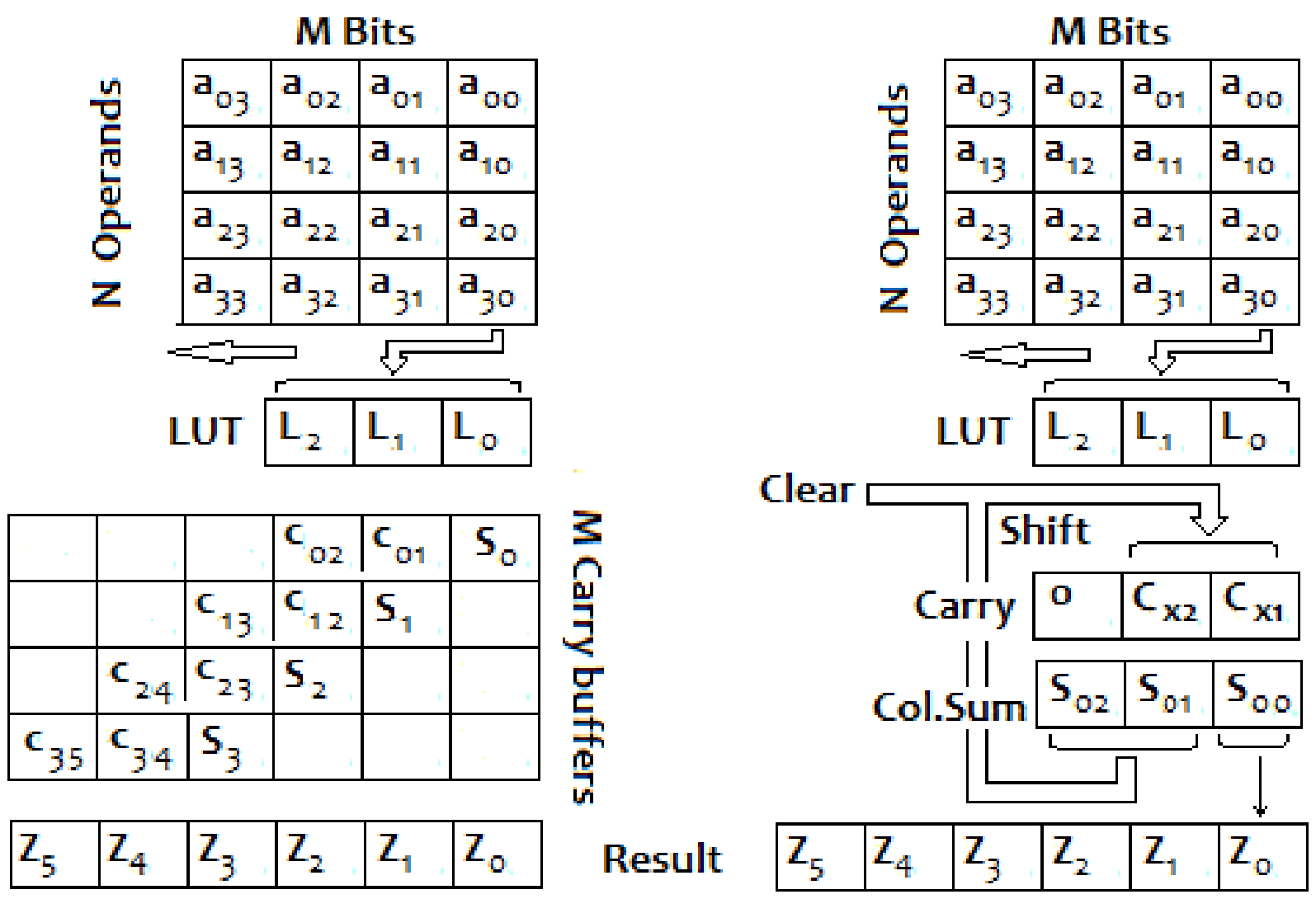

(a) (b)

**Figure 5.** 4x4 addition using (a) Algorithm-1 and (b) Algorithm-2

Figure 5(a) shows data flow for this algorithm. When input data is loaded, first column can be computed, generating the part of the data required from 2 to p+1 columns on left side. Second column can be computed now, which will give result for second column and 3 to p+2 columns to the left. The loop is executed till all $M + p$ columns are computed. Column sums of input data can be computed in a single clock. However, full column sum $S$ has to wait till the data from previous $p$ column is computed. The algorithm can compute a column addition in one clock cycle. Hence for $M$-bit serial addition, we will require at least $M + 1$ clocks.

***Algorithm-2:*** This algorithm is similar to the first one but carry generated in a column is fully added to the column on the left. The advantage here is that there is no need to store the columns of the partial sum in $p$ carry buffers. The algorithm is as follows:

a) For every $i$-th column from LSB to MSB, apply $N$-row data in the column to LUT.

b) Output of the LUT is added with carry buffer. LSB of this addition is copied to $i$-th bit of $Z$ buffer.

c) Remaining bits are shifted 1 column to the right and copied to carry buffer.

d) Increment $i$. Repeat the partial sum computation for all columns.

After all the columns are added, remaining bits in the carry buffer are copied to $Z$ register.

This algorithm also can compute $M$-bit wide $N$-operand addition in $M + 1$ clocks, but it has lower memory requirements.

### 3.3 Optimizing the Algorithms

Both algorithms can compute in *M+1* clocks. Implementing large LUTs may require large area. Hierarchical implementations with several levels of LUTs also may offer some solution. However, as the number of rows increases, the asynchronous path delay in calculating column sum may exceed one clock interval, introducing additional delay in synchronizing the output with other data paths.

Some simple optimization steps can be incorporated into the above algorithm. For example, it is possible to keep the size of carry buffer limited to $p$ bits instead of using $M + log_2(N-1)$ bits which represents the upper bound on number of bits in the result. It is also possible to merge $S$ and $Z$ buffers. Partial column sums can be grouped together and may be computed in parallel.

## 4. MULTI-OPERAND SERIAL ADDER

### 4.1 4-operand, 4-bit (4x4) Addition Operation

Figure 6 shows block diagram of a 4xM serial adder, implemented using Algorithm-2 described in section 3.2.

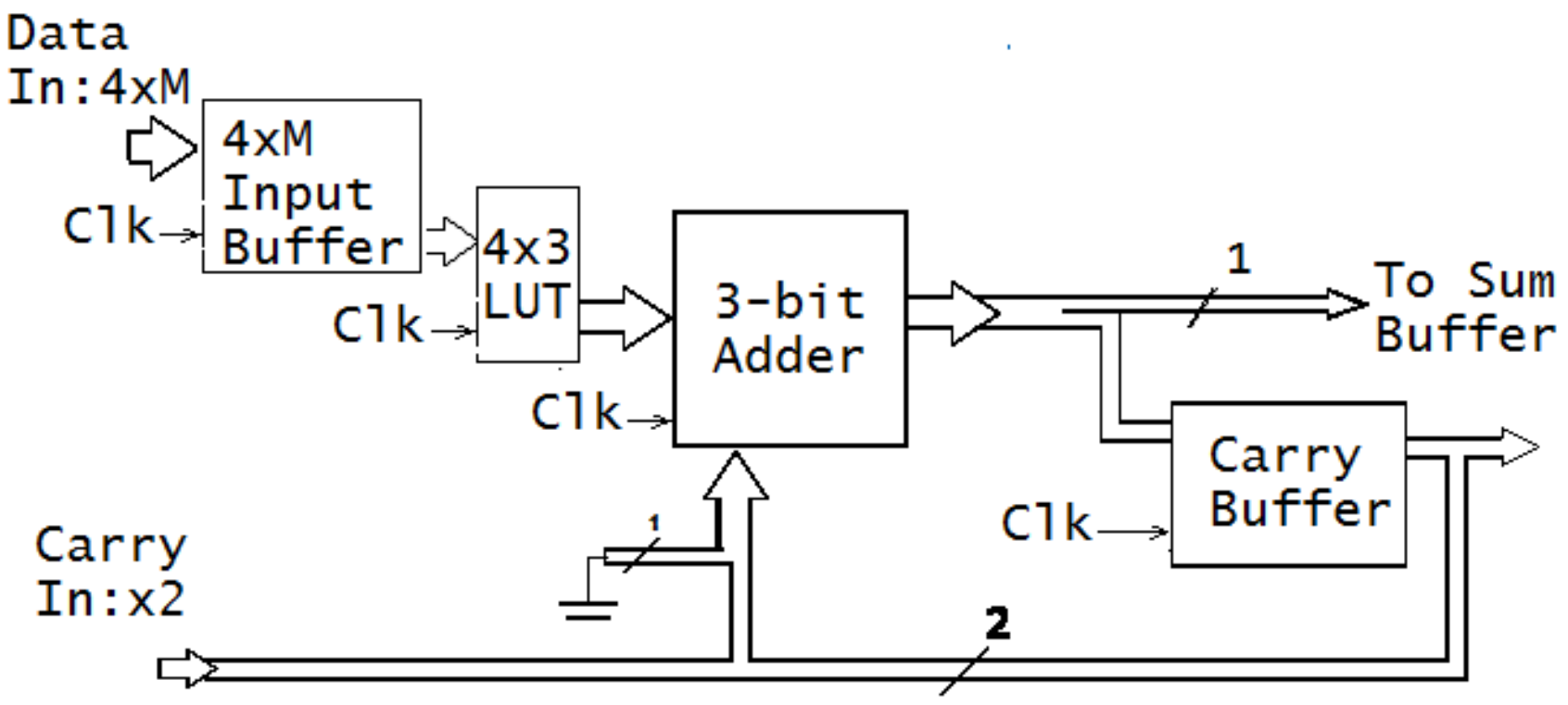

**Figure 6.** Implementation of 4xM Serial adder (Algorithm-2)

From the above theorem, maximum carry for 4-operand addition is 3, i.e., $11_{(2)}$. We prefix it with a zero to make it 3 bit wide or $011_{(2)}$. The maximum value of one column sum is $100_{(2)}$. Maximum output from the 3-bit adder is therefore $100_{(2)} + 011_{(2)} = 111_{(2)}$. Therefore, there is no overflow beyond two bits. Working of the adder shown in Figure 6 is given below:

1) Load input buffer with data to be added and clear carry buffer.
2) For column $i$ in 0 to $M-1$, apply column as input to LUT and get the 3 bit output. Add it with contents of carry buffer. Shift LSB to $i$-th bit of output buffer. Shift higher 2 bits to right and copy to carry buffer.
3) Increment $i$ and repeat 2).

Step 3) can be integrated into step 2) to reduce a clock, by transferring all three bits of 3-bit adder.

Implementation of 3-bit adder in Figure 6 requires some attention. As it is a 2-operand addition, a 3-bit full adder may be used. However, knowing that $C$ has only four possible values $(00{:}11)$ and $L$ (output of a LUT) has only 5 possible values $(000{:}100)$, only 20 out of 64 possible values will be applied as input to 3-bit adder. Therefore, full 3-bit addition will not be required at any time. It is possible to build a LUT or optimized combinatorial circuit for the 3-bit addition.

Construction of a $16x16$ adder using $4x16$ adder module is discussed in section 7. While implementing arrays of multi-operand adders, e.g. for neural networks, each $4xM$ adder may have one LUT. As only one column is added at a time, only one LUT is sufficient for the 4-operand adder. Size of LUT is fixed by choice of $N$ (4) and is independent of width of input i.e., $M$ bits. As $N$ increases, size of LUT decoder grows geometrically. Therefore while constructing adders with more than 4 operands, it is better to use 4-operand adder as a basic module. As the implementation does not require large area, use of one LUT per $4xM$ adder is a good practice.

## 5. MULTI-OPERAND PARALLEL ADDER

### 5.1 Fast Implementation of 4x4 Addition

A fast parallel implementation of 4x4 adder is given in Figure 7. It uses multiple copies of LUT given in Figure 4. Longest path has 4 LUTs and hence a delay of 16 gates. However, modularity of the design supports pipelining of the adder, allowing much faster performance. This implementation has 6 gate delays (1 for LUTs in level 1, 3 for LUTs in level 2, 1 for half adder and 1 for OR gate).

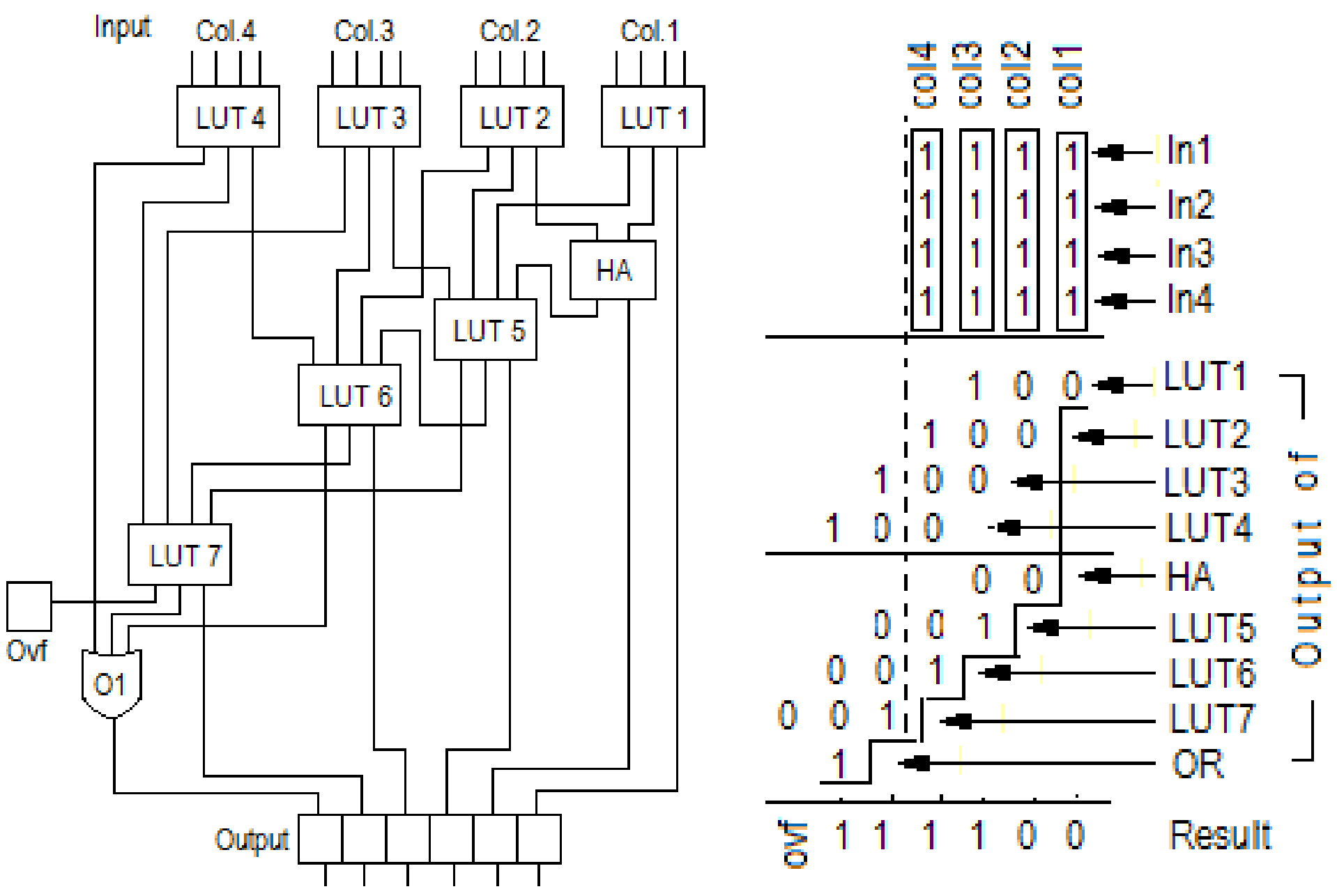

(a) (b)

**Figure 7.** Fast Implementation of 4x4 adder (a) Schematic and (b) Example

### 5.2 Implementation of Larger adders

Larger adders can be implemented by extending the 4x4 adder shown in Figure 7, to any number of operands of any size (bits). The size of such adders will increase as the number of operands increases. Another way of implementing larger adders is to use 4x4 adder as basic module as shown in Figure 8. As this is a parallel implementation, the result can be obtained within a few clock cycles.

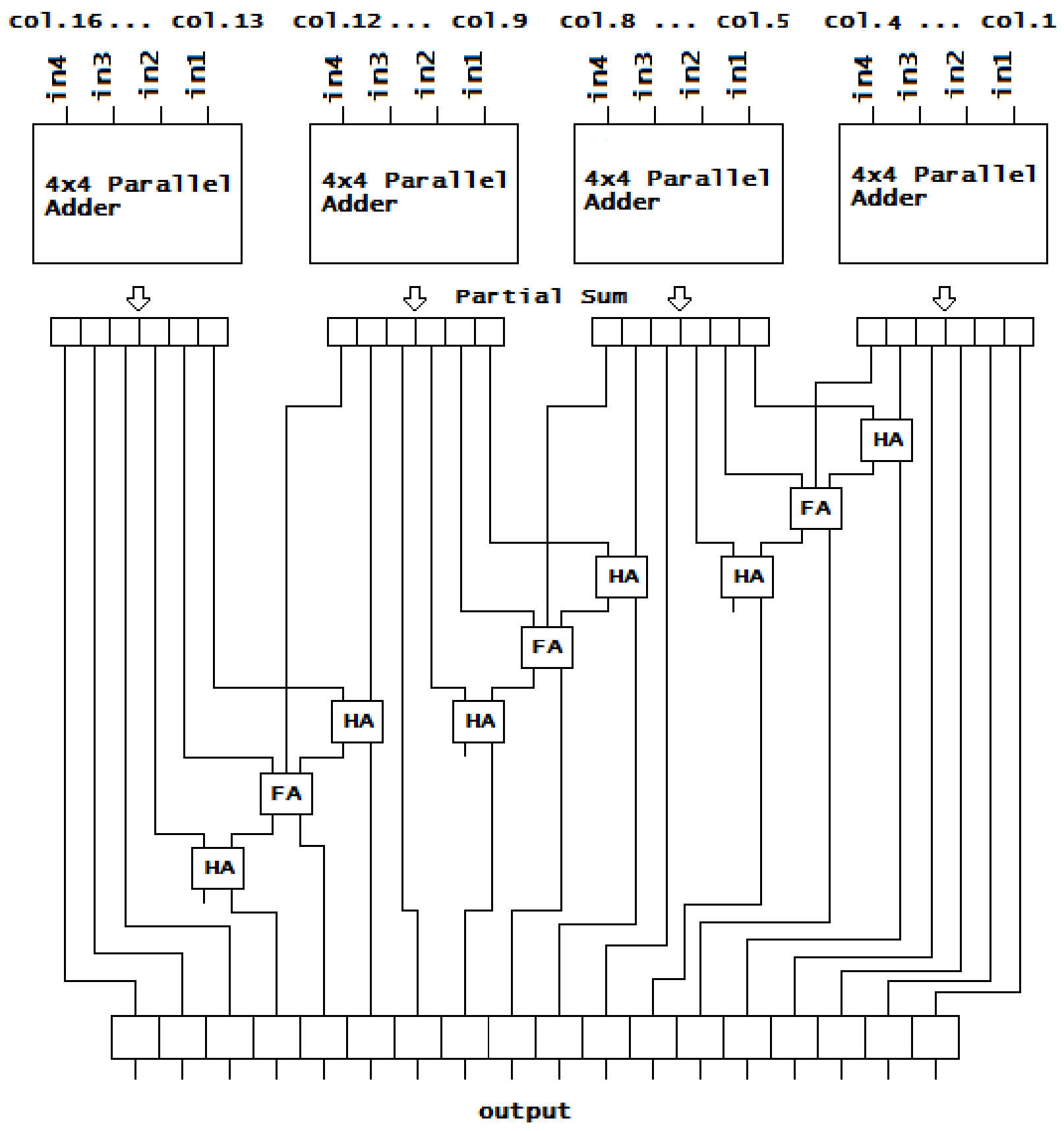

**Figure 8.** Implementation of 4x16 adder using 4x4 adder

## 6. SERIAL VS PARALLEL IMPLEMENTATIONS

Almost every arithmetic computation can be implemented in parallel as well as in serial mode. The number of serial units that can be realized in a given area is higher than the number of parallel units. However, parallel units compute faster. Assume that the parallel unit computes a result in 1 clock but the serial unit needs 10 clocks. If the area taken by parallel units is 15 times that of serial units, it should be possible to implement 15 serial units in the same area as one parallel unit (assuming no overheads). Given 10 clocks, the serial units will perform 15 computations while the single parallel unit can complete only 10 computations. Therefore, the throughput of a set of serial implementations can exceed the throughput of parallel implementation. This argument assumes that there are enough computations to keep the units busy. Such scenario exists in a massively parallel environment like that of a DL system. An observation is made here on the trade-off between implementing serial and parallel units.

***Lemma 3:*** *In a massively parallel environment, the throughput of a set of serial execution units can exceed the throughput of parallel units occupying the same area as the set of serial units if the ratio of areas is greater than the ratio of execution times.*

***Proof:*** In a massively parallel environment, number of pending operations will be in excess of available resources. Let $A_s$ and $A_p$ represent the areas of serial and parallel units. The parallel units take a larger area and hence the ratio $\frac{A_p}{A_s} = R_A > 1$. Similarly, let $T_s$ and $T_p$represent the number of clocks required to execute serial and parallel units. As parallel units execute faster, the ratio of time taken to execute one instruction is $\frac{T_s}{T_p} R_T > 1$. Considering area of one parallel unit, the number of serial units that can fit into the same area is $R_A$. For a given amount of time T, parallel unit can execute T/Tp instructions. Similarly the set of serial units can execute $R_A \left(\frac{T}{T_s}\right)$ instructions. Now, if the throughput from the set of serial units is larger than that of a parallel unit, we can write

$$R_A \left(\frac{T}{T_s}\right) > \left(\frac{T}{T_p}\right), \text{ or } R_A > \left(\frac{T_s}{T_p}\right), \text{ or}$$

$$R_A > R_T \qquad \text{QED}$$

The interconnection and other overheads are ignored in the lemma as they can be absorbed into the area parameters. Some of the large modules like multiplier do satisfy this condition and hence it is better to implement more number of serial units than parallel units. Many open source soft core processors provide options to implement serial or parallel multipliers, which will find this lemma useful. For example, HF-RISC, a soft core RISC processor based on MIPS I architecture provides options for parallel, serial and software multipliers to suit the area/cost demands [2016 Johann]. Comparative performance of serial and parallel implementations of a hypothetical module is presented in Figure 9. The speed ratio $R_T$ is assumed to be 17:1. The figure shows the number of operations performed (represented as throughput) for two area ratios, viz., $R_A$=12 and $R_A$=20. It can be easily seen that when the area ratio exceeds speed ratio, serial units perform better.

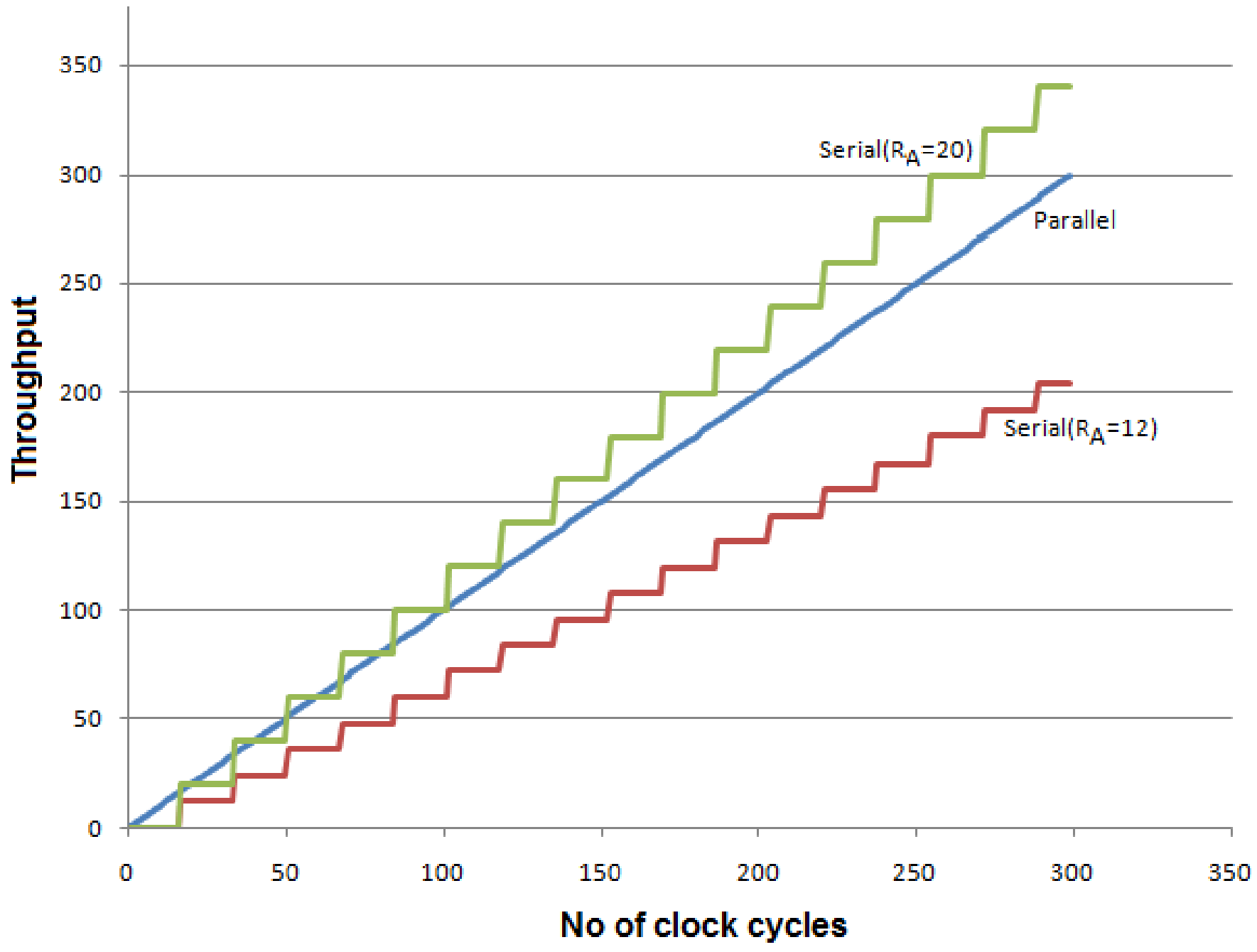

**Figure 9.** Performance of Serial and Parallel modules in massively parallel environment with Serial to Parallel area ratio of 1:12 and 1:20 and speed ratio of 17:1.

### 7. GENERALIZED ALGORITHM FOR RECONFIGURING ADDER MODULES

The number of operands in a neural network varies depending on several parameters like the learning algorithm, type of the input, end application, etc. Therefore, reconfiguring the existing hardware becomes necessary. In the following section, an algorithm to implement larger modules using 4-operand adder as a basic unit is explained (see Table 4). A set of $4xM$ adders can be used to construct a $16x16$ adder. Overall procedure will be as follows (See Figure 10):

i) Divide the operands in to 4 groups of 4 operands each and input each group to a $4x16$ adder. Separate the sum and carry parts to obtain four sum values S3:S0 and carry values C3:C0.

ii) Add four sum values S3 to S0 to generate the final sum S and carry C4.

iii) Add all carry values C4 to C0, to generate final carry C.

iv) Concatenate {C, S} to get the final output.

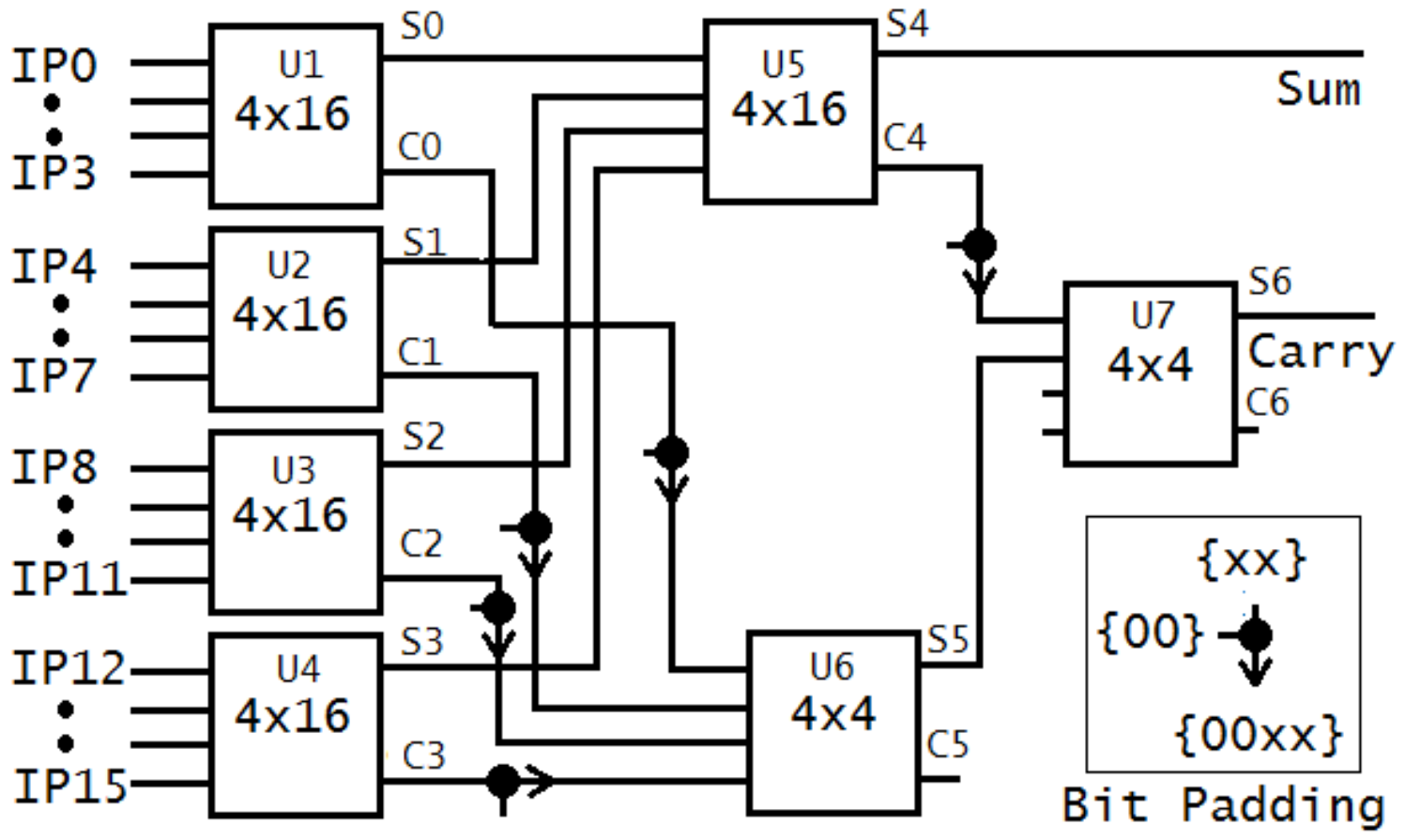

**Figure 10.** Modular implementation of 16x16 adder

Few details in Figure 10 need attention. Carry output from U1 to U5 are zero padded to match the $4x4$ adders U6 and U7. S5 from U6 has an upper bound of 12, i.e., $1100_{(2)}$. So there is no carry from U6 i.e., C5 = 0, always. On the other hand, carry C4 has an upper bound of 3, i.e., $0011_{(2)}$ as U5 has only 4 inputs. Therefore, sum S6 from U7, will have a maximum value of 15, i.e., $1111_{(2)}$. Therefore, there is no carry from U7, i.e., C6 = 0. This is as expected from Theorem – maximum carry for 16-operand addition is 15. It is possible to replace U6 by a 4x2 adder. In that case, S5 and C5 have to be concatenated {C5, S5} and apply it as input to U7. Further, as U7 has only two inputs, a 2x4 adder is sufficient. Carry C6 will be only one bit but may be ignored as it will always be zero. We have used $4x4$ adder only for modularity of implementation.

Using this example as a template, it is also possible to generalize the implementation of $N$-operand adder using $4xM$ adder as a basic module. Select $P = N/4$ adders. Generate $S$ & $C$ for each of $P$ adders. Sum $S$ from each of the $P$ adders have to be added together till a single $S$ term is obtained. Each such addition will generate additional $C$ terms. Add all $C$ terms and concatenate with the $S$ term to obtain the final sum.

**Table 4.** Algorithm for reconfiguring adder modules

```
Assume 4x16 and 4x4 adder modules are available for
reconfiguration.

Objective
   To configure a 16 bit adder with N operands

Structure
   Module Add4x16(in[4][16], s, c)
   Module Add4x4  (in[4][4], s, c)

   Configuration {
      Add4x16 A[1...L][1...R]
      Add4x4  C[1...L][1...R]
      Add4x4  B[1][1]
   }
Algorithm
   N1 = N
   L = ceil(log(N1)/log(4))
   For i = 1 to L
      R = N1/4
      For j = 1 to R
         For k = 1 to 4
            p = (j-1)*4 +k
            //Place sum Adders
            If (i == 1)
               //Adders
               A[i][j][k] = in[p]
            Else if (i == 2)
               A[i][j][k] = A[i-1][p].S
            //Place carry adders
            If (i >= 2)
               C[i][j][k]= A[i-1][p].S
            If (i >= 3)
               C[i][j+R][k] = C[i-1][p].S
      N1=R
   B[1][1] = A[L][1].C
   B[1][2] = C[L][1].S
```

## 8. MULTI-OPERAND ADDERS IN ARTIFICIAL NEURONAL IMPLEMENTATION

Many DNNs require multi-operand adders as one of the fundamental modules. For example, in an Auto Resonance Network [2020 Aparanji et al.], every node has as many resonators as the number of inputs. The output of every resonator is added and scaled to compute the output of ARN node. So, the number of resonators available to the node is the number of inputs supported by the ARN node.

$$y = \frac{4}{Nk^2} \sum_{i=1}^{N} X_i \,(k - X_i) \qquad (21)$$

However, if more inputs are required, the partial outputs from adder need to be temporarily stored. These can be added later after all inputs are processed. In the current implementation, addition is performed using a multi-operand adder. This adder can be configured to match the actual number of inputs to save computational time. A regularization unit at the output of ARN nodes may be used to normalize the outputs to be in the range as required. As an illustration, an image recognition system using a two-layer ARN is considered for the implementation [2019 Mayannavar et al.]. It has 16-inputs to the resonators and the output of all the resonators are added using a 16x16 adder.

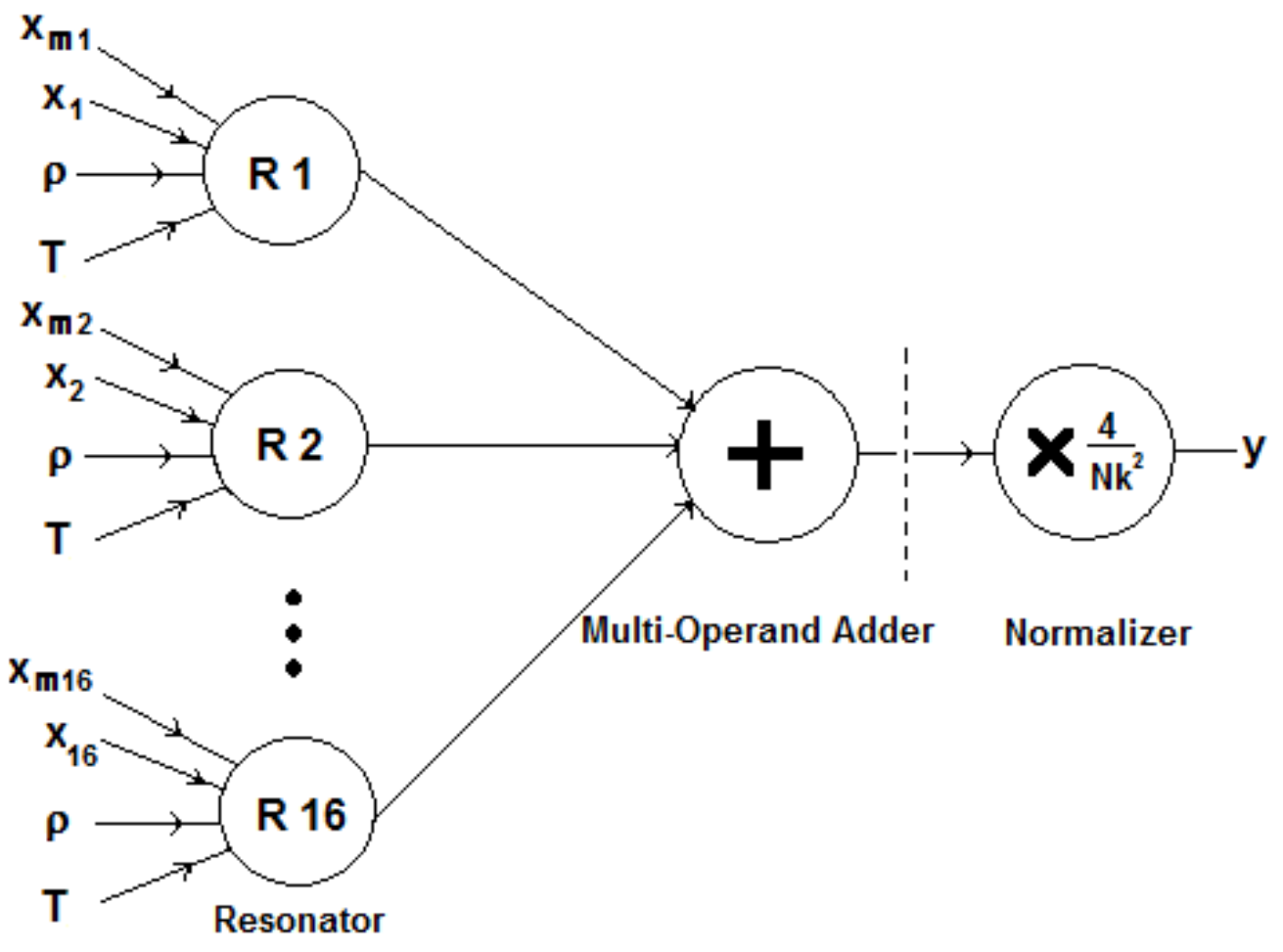

**Figure 11.** Structure of image recognition using ARN with 16-inputs

The node with 16 inputs followed by resonators, multi-operand adder is shown in Figure 11. The inputs to the resonator are x, $x_m$ and control parameters ($\rho$ and T). The output of a node with N inputs is calculated using the formula given in eqn. (21).

Similar concept of implementing ARN node may be extended to other neural architectures like MLP (Multi-Layer Perceptron). A typical neuron in MLP will have set of inputs corresponding to the strength of synaptic connections. Each of these inputs are multiplied with their synaptic weights and added together followed by a non-linear activation function. Therefore, a structure of an 16-input perceptron will contain set of multipliers, multi-operand adder and an activation function. In addition to these basic modules it can include several other modules depending on its end application, input type, learning algorithm and many other parameters. The structure of 16-input perceptron shown in Figure 12 has sixteen serial multipliers and multi-operand adder followed by an activation function. The simulation results of 16-input neuron is shown in section 9.

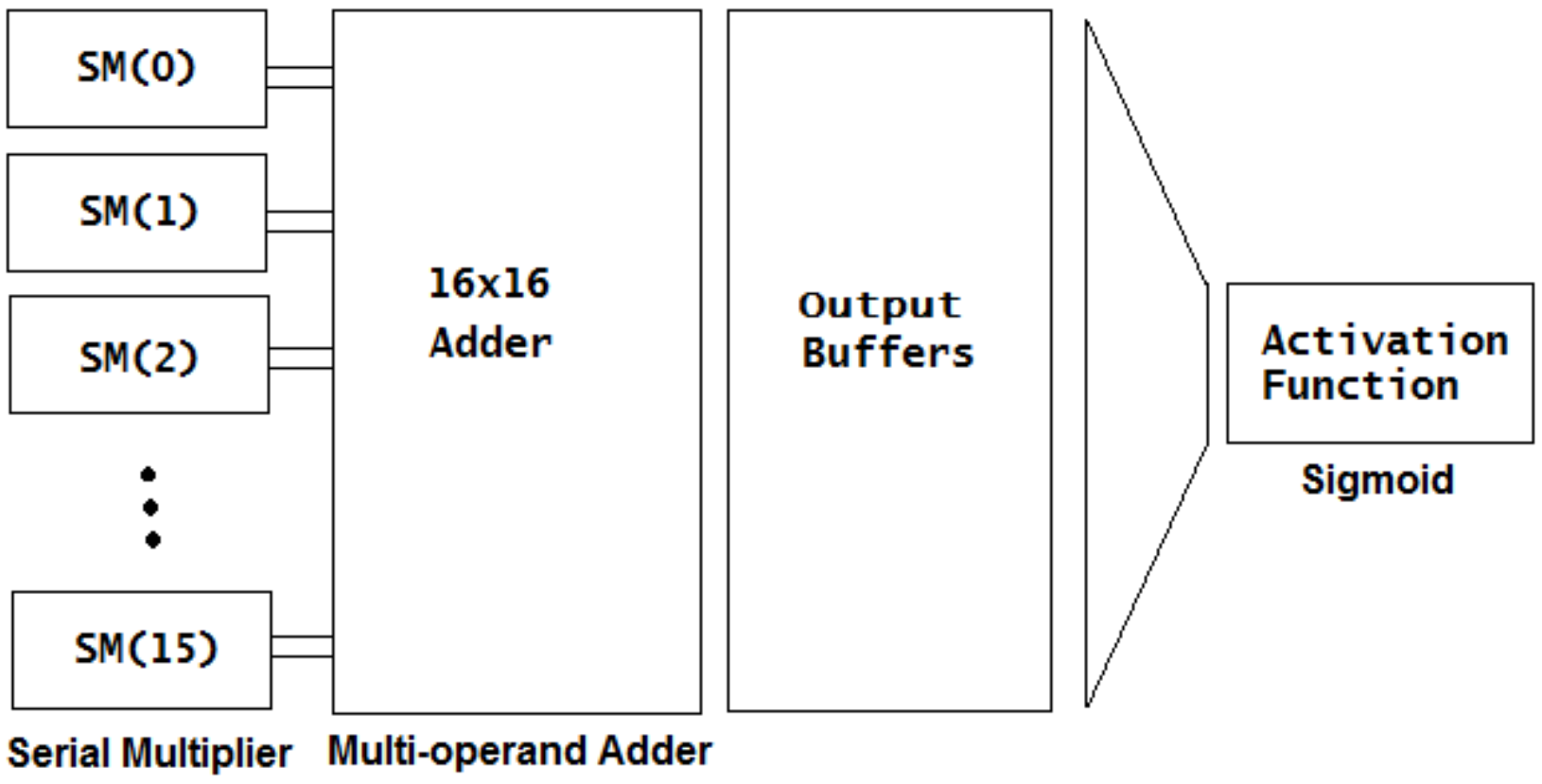

**Figure 12.** Structure of a16-input perceptron in MLP

## 9. SIMULATION RESULTS

### 9.1 4x4 Serial Adder

The simulation result of 4x4 serial addition using single LUT is shown in Figure 12, which shows the addition of four numbers in base 16 i.e., A+F+1+2=1C$_{(h)}$. From the above theorem and corollary, 2-bits are sufficient to represent the carry of 4x4 addition. A total of 6-bits are sufficient to represent the final sum of 4x4 addition. Output of LUT for four columns is {2, 3, 1, 2}. Four columns will require four clock cycles. Stable data is available at the fifth clock.

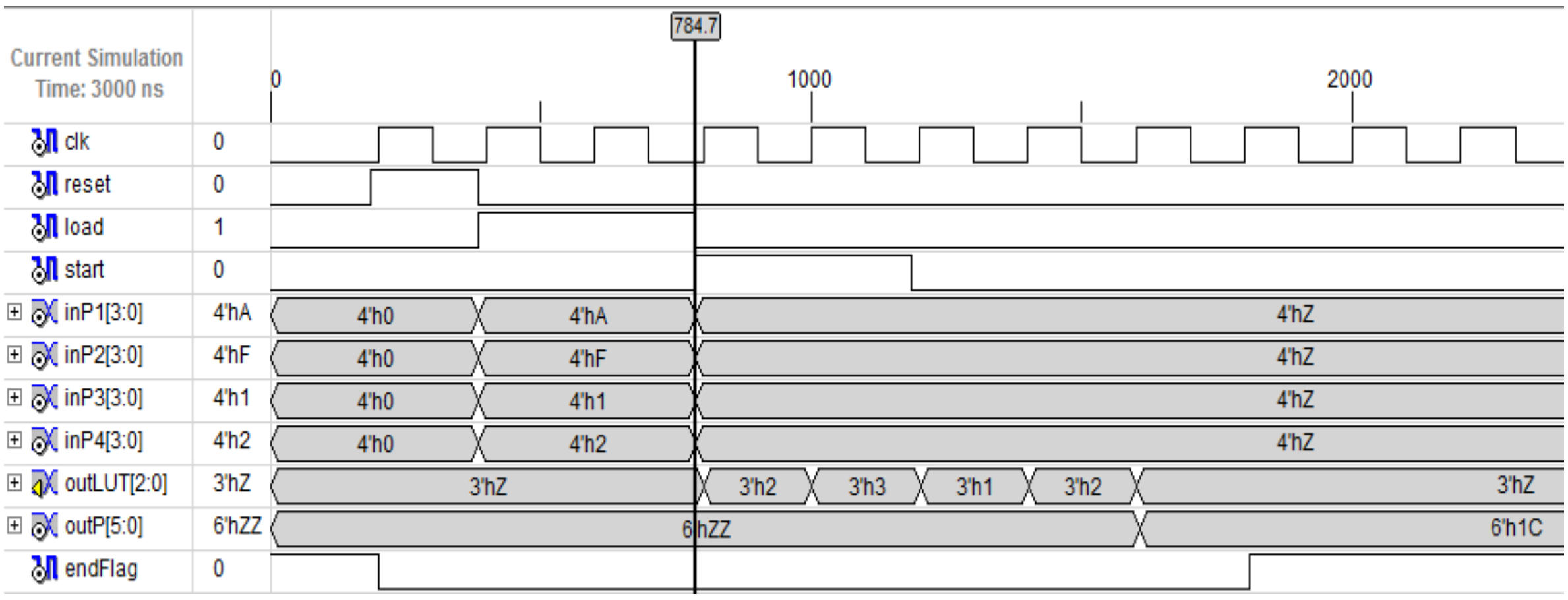

**Figure 12.** Simulation result of 4x4 serial addition

### 9.2 4x4 Parallel Adder

The simulation result of 4x4 parallel addition is shown in Figure 13. The sum of four columns is the output of LUTs. It can be noticed that the output of all the LUTs {2, 3, 1, 2} are available in parallel and therefore, total computation time is reduced and the output is available within a single clock.

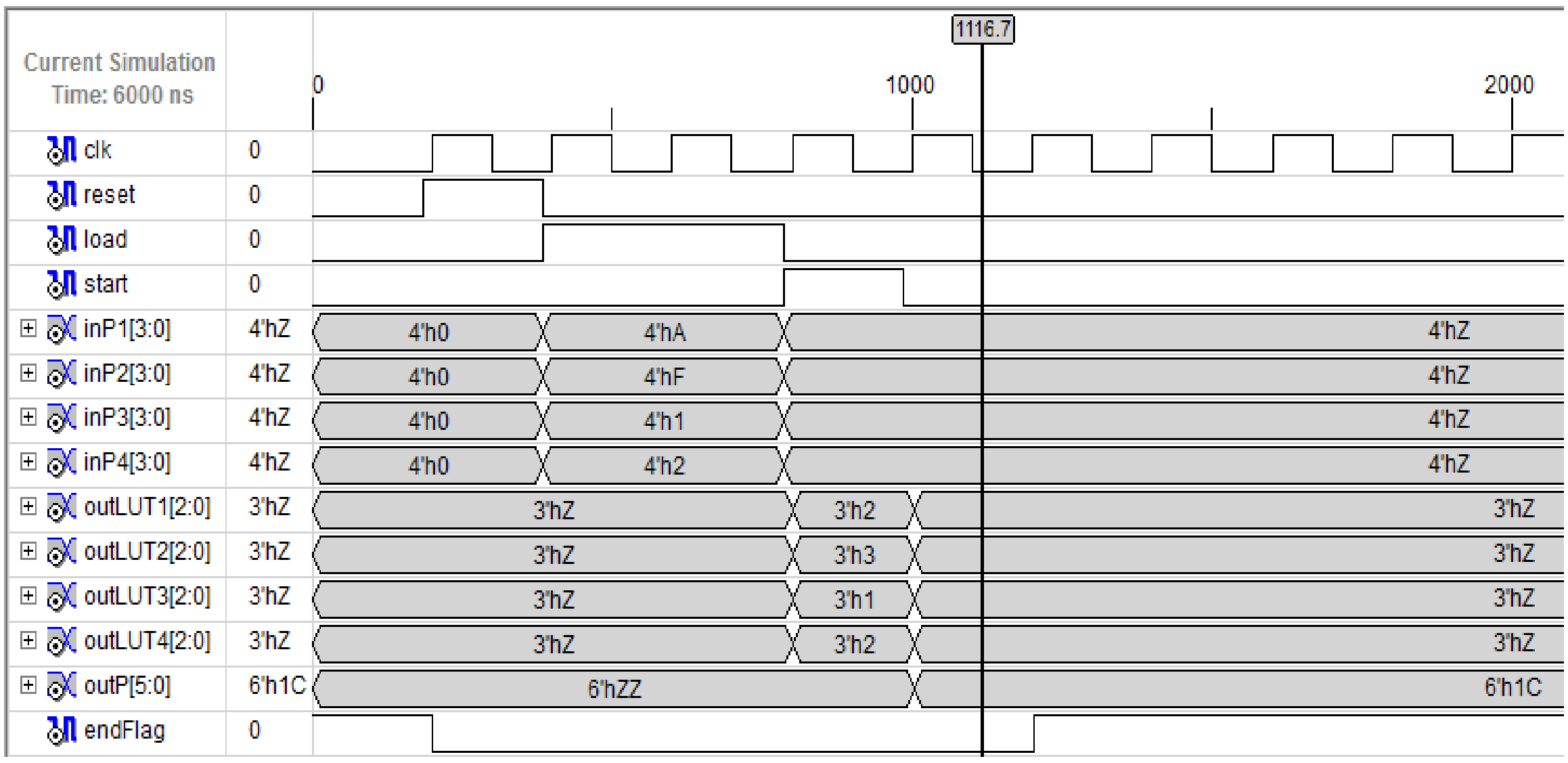

**Figure 13.** Simulation result of 4x4 parallel addition

### 9.3 4x16 Serial Adder

The 4x4 serial adder module can be easily extended to 4x16 adder simply by increasing the number of iterations to 16. Correspondingly we will perform 16 column additions. The simulation result of 4x16 addition is shown in Figure 14, it shows the addition of four 16-bit numbers in base 16 i.e., A234+FFFF+0A2D+FF7F=$2ABDF_{(h)}$. According to the theorem, a total of 18-bits are sufficient to represent the final sum of 4x16 addition. The operation can be completed within 16+1 clock cycles.

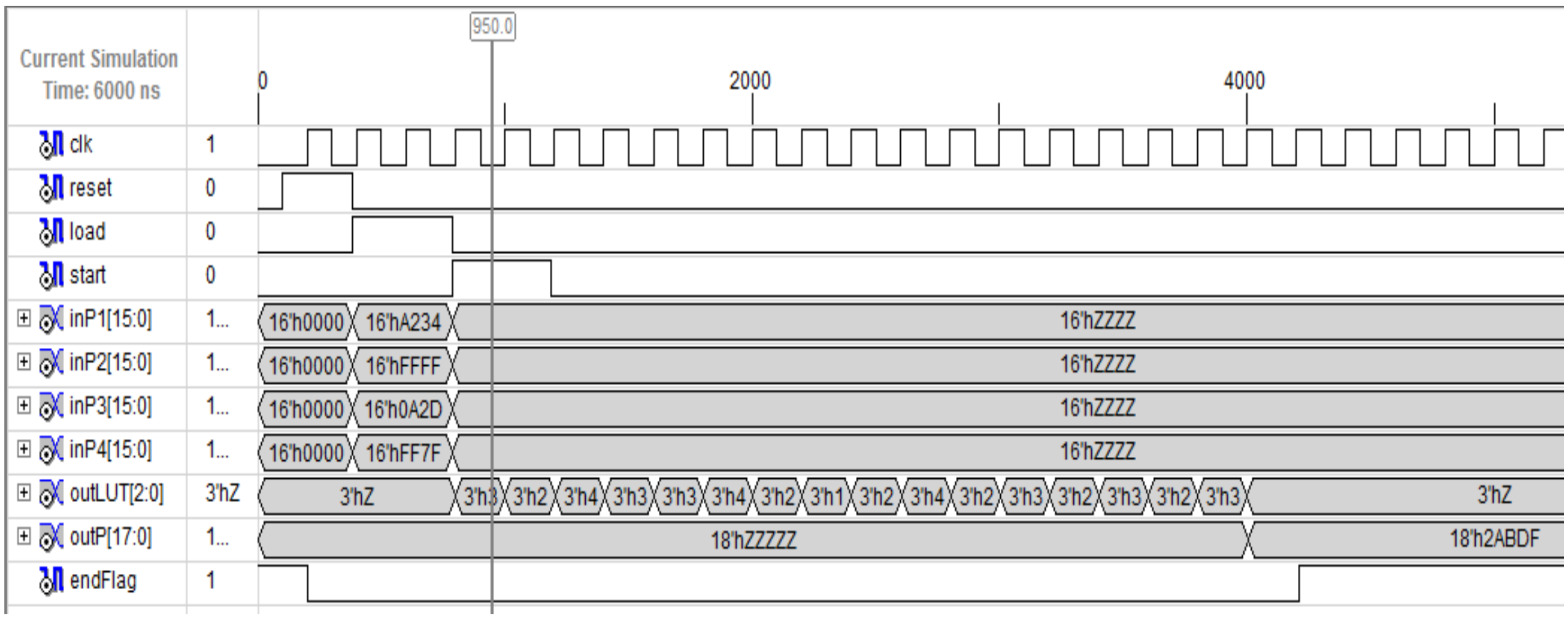

**Figure 14.** Simulation result of 4x16 serial addition using single LUT

### 9.4 16x16 Adder

Using a generalized algorithm described in section 7, 16x16 adder is implemented using 4x16 adders. The simulation result of 16x16 adder is shown in Figure 15.

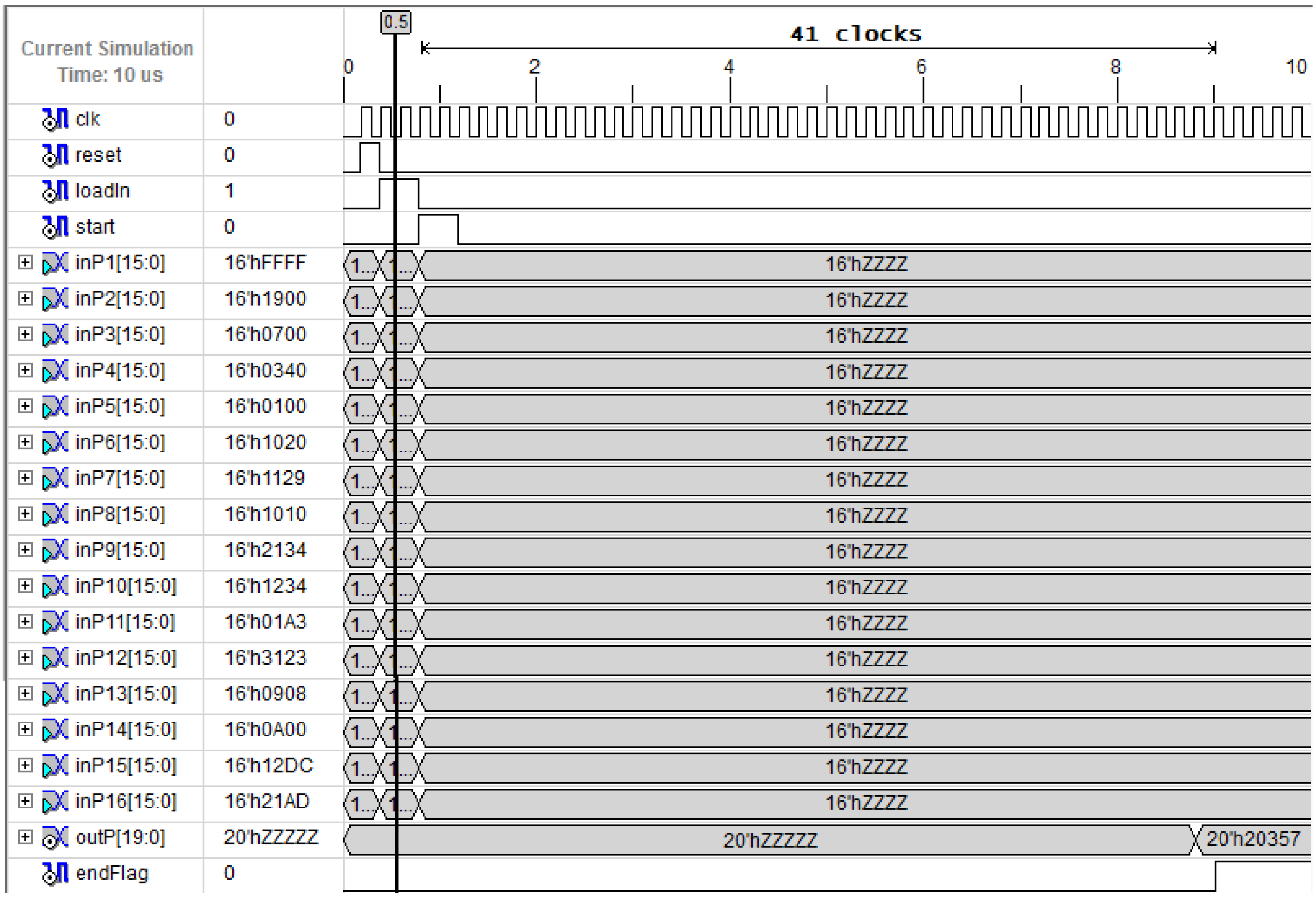

**Figure 15.** Simulation result of 16x16 addition

## 10. PERFORMANCE COMPARISON OF LUT BASED ADDER WITH CONVENTIONAL ADDER

The effect of massive parallelism on the performance of area constrained processing element arrays may be seen in the following chart (see Figure 16).

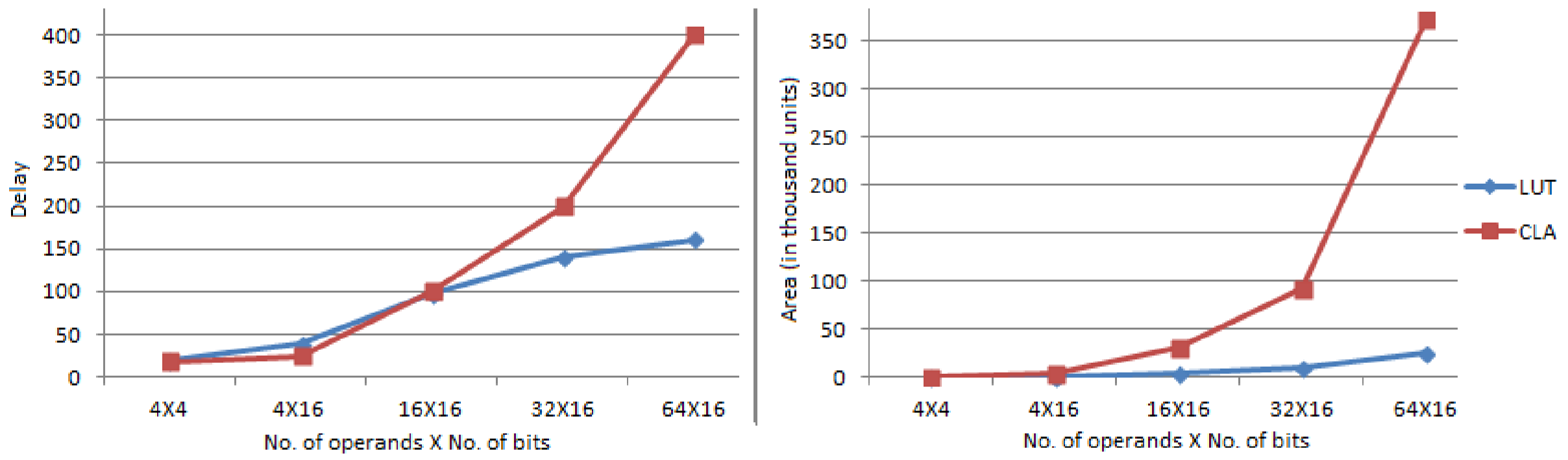

(a) (b)

**Figure 16.** Effect of massive parallelism on Gate Delay and Gate Area of CLA and LUT based adder

The delay and area are considered in units of (two input) gates. The Carry Look Ahead (CLA) adder is compared with the LUT based four-operand adder. LUT for 1-bit addition, shown in figure 4 has a longest path of 4 gates and overall area of 25 gates. The gate delay and gate area of two-operand 4-bit CLA are 9 and 50 respectively [2013 Jovanovic]. These basic units are further extended to calculate the gate delay and gate area for multi-operand addition. Figure 16 (a) shows the **gate delay** for CLA and proposed LUT based adders. It can be seen that, LUT based adder is slower when the number of operands is less than four. However, when the number of operands is large (say 16) LUT based adder is faster than the CLA.

The figure 16 (b) shows the **gate area** for CLA and LUT based adder. The area taken by both the adders are almost same when the number of operand is less than four, but when the number of operands increase, LUT based adder takes very less area as compared to CLA. Therefore, LUT based adders have both area and speed advantage in a massively parallel computing environment.

Performance comparison of LUT based adder and CLA is carried out for different number of operands and bit width (see Figure 17). It can be noticed that, when the number of operands and the number of operations are less (say 4), performance of both the adders is almost similar. However, when the number of operations increases, CLA performs better as shown in Figure 17(a). Increasing the bit width also shows similar trend: performance of both adders is same when the number of operations are less but as the number of operations increase, LUT based adder performs significantly better (see Figure 17(b)). Same holds good for large number of operands, number of operations as shown in Figure 17(c) and (d).

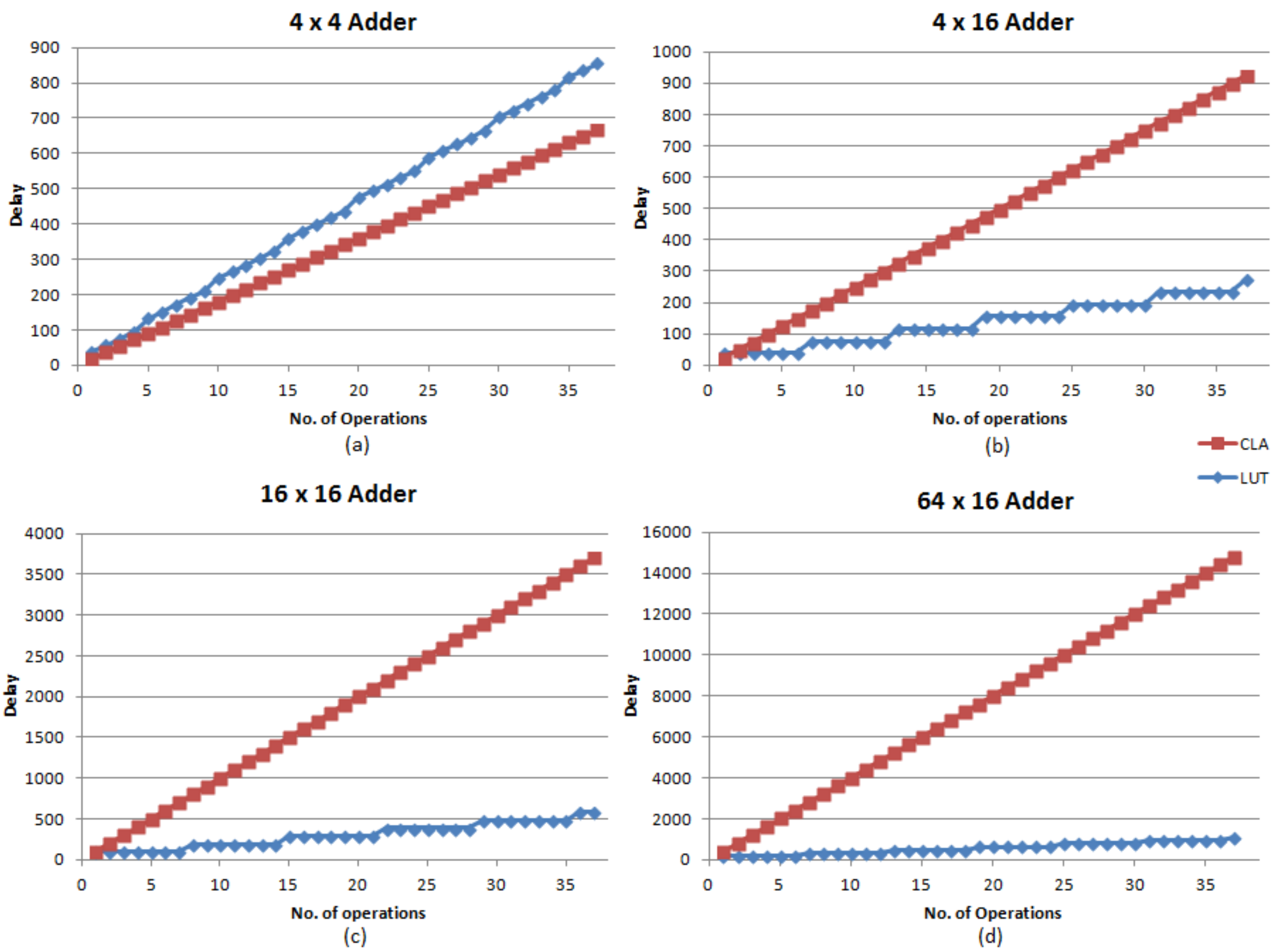

**Figure 17. Performance of LU based and CLA adders for different adder size**

LUT based adders have performance advantage when the number of operations (ops) are large. The graph shown in Figure 18 depicts such advantage for different adder complexity. The performance advantage is calculated using the eqn. (22)

$$PerformanceAdvantage = \frac{d_g(CLA)}{d_g(LUT)} \quad (22)$$

Where $d_g$ represents the gate delay considering the longest path. It can be seen from the figure 18 that CLA has performance advantage when the adder complexity and number of operations are less. However, the LUT based adder has performance advantage when the number of operands crosses 4. In general, we can see that LUT is faster if performance advantage is greater than 1 and CLA is faster otherwise.

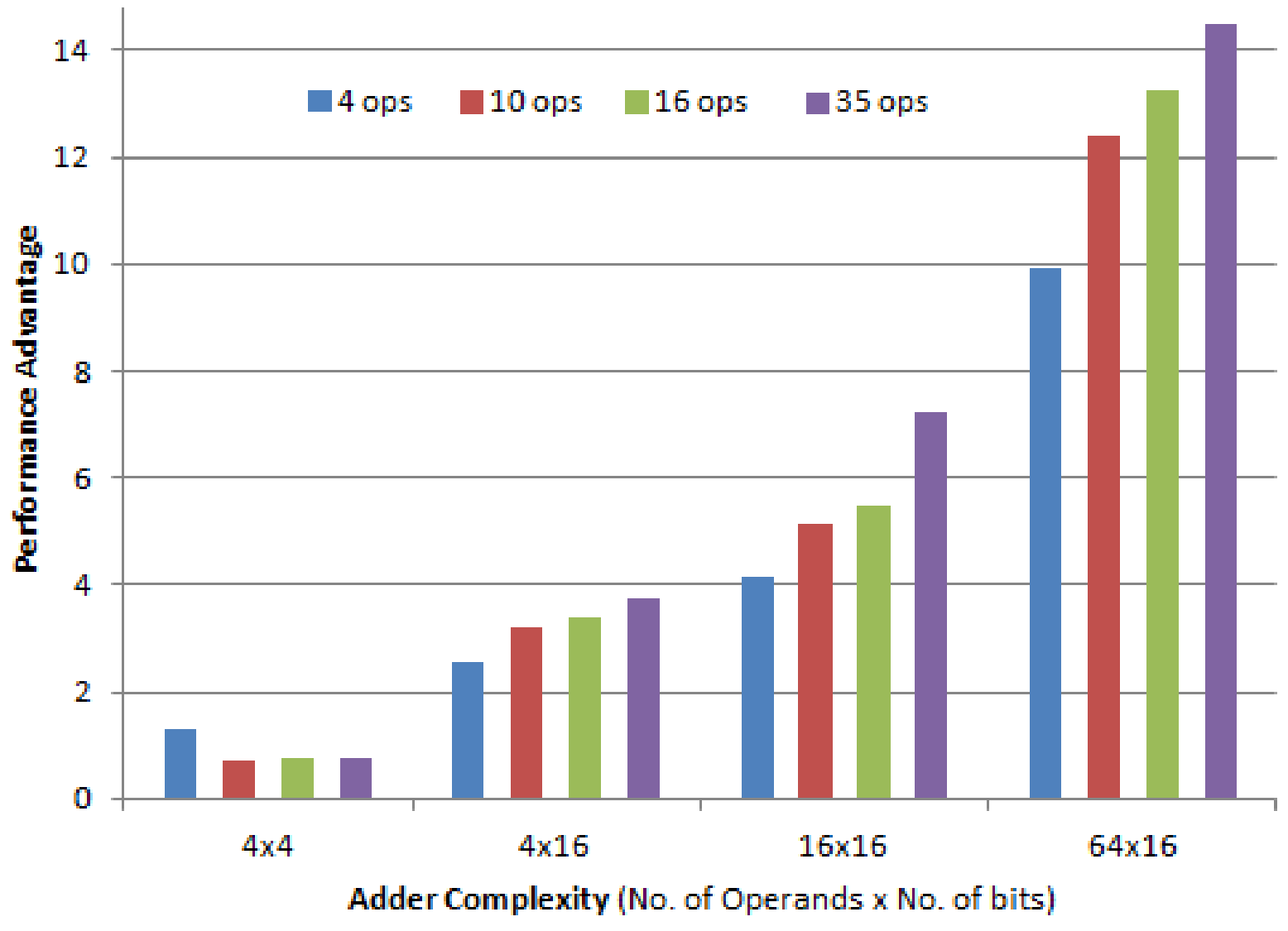

**Figure 18.** Performance advantage of LUT based adders

**11. CONCLUSION**

The paper discussed the need for fine grain re-configurable modules in processor design. Contemporary designs have to focus on the evolving DLNN architectures and applications. These computing scenarios involve use of massively parallel processing, which represents a different kind of computational load on the processors. Most of the neural computing architectures require a multi-operand adder at the axonal output as the number of inputs to a neuron can be very large. The paper presents a new approach to design of multi-operand adders, instead of using chained two operand adders. The paper discusses all the necessary maths to implement such adders. Required Lemmas have been presented along with their proofs. As the number of operands changes, the number of carry bits also changes. Equations to quickly calculate the number of carry bits have been presented. A 4 bit 4 operand adder was developed as a basic module. Arrays of this module may be reconfigured to implement larger adders. Performance of serial adders seems to improve over that of parallel adders for a given area. Simulation results of all the modules have been presented.

## Acknowledgment

We thank Prof. Aviral Srivastava of Arizona State Universtity for providing contiuous feedback and advice. We also thank Prof Pradip K Das of IIT Guwahati for the encouragement provided. The authors would like to thank SGBIT Belagavi and C-Quad Computers, Desur IT Park, Belagavi and for all the facilities and support provided.

## Author Biographies

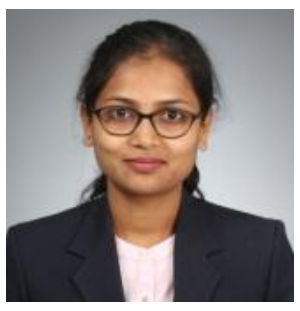

**Shilpa Mayannavar** is an Assistant Professor, Dept. of ECE, S G Balekundri Institute of Technology Belagavi. She is also a Research Scholar at C-Quad Research, Belagavi, Karnataka, India. She has obtained Bachelor of Engineering in Electronics and Communication Engg., (2012), Master of Technology in VLSI Design and Embedded System (2014) and Ph.D in Processor Design for Neural Networks (2020) from Visvesvaraya Technological University (VTU), Belagavi. Her research interests are Processor design, Artificial Intelligence and Neural Networks and Neural network Hardware.

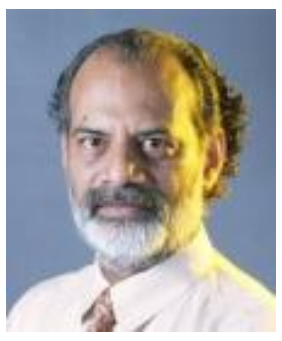

**Uday Wali** is a Professor in Dept. of ECE at S G Balekundri Institute of Technology, Belagavi, Karnataka India. He has obtained Bachelor of Engineering in Electrical and Electronics Engineering from Karnataka University Dharwad (1981) and Ph.D from IIT Kharagpur (1986). He is a CEO of C-Quad Computers, Desur IT Park, Belagavi. His research areas of interest are Artificial Intelligence, Neural Networks, Cognitive Radio, Processor Design and etc.